\documentclass[11pt]{article}
\pdfoutput=1
\usepackage[utf8]{inputenc}
\usepackage{multirow}
\usepackage{amsmath, amsfonts, amssymb}
\usepackage{comment}
\usepackage{graphicx}
\usepackage{psfrag}
\usepackage{amsthm}
\usepackage[usenames,dvipsnames,svgnames,x11names,table]{xcolor}
\usepackage{enumerate}
\usepackage{arydshln}
\usepackage{soul}
 \usepackage{slashed}
\usepackage{subcaption}
\usepackage{mathrsfs}
\usepackage{ytableau}
 \usepackage{a4wide}
  \usepackage{tikz}
  \usepackage{tikz-cd}
  \usetikzlibrary{shapes.geometric}
  \usepackage{tcolorbox}
  \usepackage[usenames, dvipsnames]{xcolor}
  \definecolor{dark-gray}{gray}{0.20}
  \definecolor{gray}{gray}{0.30}
  \definecolor{light-gray}{gray}{0.80}
  \definecolor{dark-red}{rgb}{0.7,0,0}
  \definecolor{dark-green}{rgb}{0.1,0.4,0}
  \definecolor{dark-blue}{rgb}{0.3,0.3,0.7}
  \definecolor{light-blue}{rgb}{0.8,0.8,1}
      \definecolor{swamp}{RGB}{240, 199, 197}
      
  \usepackage{pifont}

\usepackage{newunicodechar} 
\usepackage{setspace}

\newcommand{\ch}{\mathrm{ch}}

\newcommand{\be}{\begin{equation}}
\newcommand{\ee}{\end{equation}}
\newcommand{\eq}[1]{(\ref{#1})}

\def\be{\begin{equation}}
\def\ee{\end{equation}}
\def\bea{\begin{eqnarray}}
\def\eea{\end{eqnarray}}

\AtBeginDocument{

}

\numberwithin{equation}{section}

\usepackage{jheppub}
\usepackage{hyperref}
\usepackage{cleveref}

\hypersetup{
	colorlinks=true,
	linkcolor=dark-blue,
	citecolor=dark-red,
	urlcolor=dark-green,
	linktoc=page
}

\theoremstyle{definition}

\theoremstyle{remark}

\crefname{appendix}{Appendix}{Appendices}

\title{\centering M-theory boundaries beyond supersymmetry}

\author{Miguel Montero$^1$} \affiliation{$^1$ Instituto de F\'{i}sica Te\'{o}rica IFT-UAM/CSIC,
C/ Nicol\'{a}s Cabrera 13-15, Campus de Cantoblanco, 28049 Madrid, Spain}
\author{and Luis Zapata$^{2,3}$}
\affiliation{$^2$ Departamento de F\'{i}sica, Centro de Investigaci\'{o}n y de Estudios Avanzados del IPN, PO Box 14-740, C.P. 07000, Ciudad de M\'{e}xico, Mexico}
\affiliation{\centering $^3$New York University Abu Dhabi, Saadiyat Island, Abu Dhabi, PO Box 128199, UAE.}
\emailAdd{miguel.montero@csic.es}
\emailAdd{lzapata@fis.cinvestav.mx}

\abstract{The chiral worldvolume theory of an M-theory boundary (the so-called M9 brane) is uniquely determined by supersymmetry and anomaly inflow. In this brief note we investigate whether alternative chiral boundary field contents  may be allowed by anomaly cancellation once supersymmetry is dropped. Even then, anomaly inflow places stringent constraints on the gauge group $G$ and matter content of the boundary worldvolume theory, which we determine explicitly. We find the most general solution to these constraints in the case where all matter fields are of the same chirality, for all simple Lie algebras except  $\mathfrak{sp}_2$, $\mathfrak{su}_{n\leq5}$, and $\mathfrak{so}_{n}$ with $7\leq n\leq 12$, and find no solutions other than the supersymmetric $E_8$ boundary of Ho\v{r}ava and Witten. However, when we extend our search to allow for any chirality in the matter fields, we find one minimal solution with gauge group $G_2$, charged matter in the $\mathsf{14}$, $\mathsf{27}$ and $\mathsf{77}$ representations, which satisfies all constraints in a non-trivial way. Therefore, it could in principle describe the low-energy theory of a novel nonsupersymmetric M-theory boundary condition, different from the Ho\v{r}ava-Witten proposal. We briefly discuss some consequences if this was indeed the case, such as the existence of a non-supersymmetric, exotic  ``$G_2$-string'' CFT in 6d, and a novel, non-perturbative, heterotic-like 10d string with gauge group $G_2\times G_2$.}

\begin{document}
\emergencystretch 3em
\hypersetup{pageanchor=false}
\makeatletter
\let\old@fpheader\@fpheader
\preprint{IFT-25-033}

\makeatother

\maketitle

\hypersetup{
    pdftitle={},
    pdfauthor={},
    pdfsubject={}
}

\newcommand{\remove}[1]{\textcolor{red}{\sout{#1}}}

\section{Introduction}
One of the aspects of M-theory that played a crucial role in the second superstring revolution was discovering that the theory makes sense on manifolds with boundary \cite{Witten:1996md}. Concretely, M-theory compactified on an interval (the so-called Ho\v{r}ava-Witten theory \cite{Horava:1995qa, Horava:1996ma}) is described by 10d $\mathcal{N}=1$ heterotic supergravity with gauge group $E_8\times E_8$ at low energies, and provides the bridge between M and heterotic string theories. The two $E_8$ factors appear, in the M-theory picture, because there is an $E_8$ $\mathcal{N}=1$ gauge multiplet living in each boundary of the interval (see \cite{Diaconescu:2003bm, Freed:2004yc} for a more systematic study of the well-definiteness of M-theory and its quantum theory on a manifold with and without boundary using this connection with $E_8$ gauge theory,  and \cite{Freed:2019sco} for a more recent study of M-theory on unorientable manifolds.) 

As was shown in \cite{Horava:1995qa, Horava:1996ma}, the precise matter content of an M-theory boundary (also known as the M-theory end-of-the-world brane, or M9 brane) is completely determined by supersymmetry and anomaly inflow from the bulk. The anomaly inflow problem is, in fact, very much over-constrained, and so the existence of a boundary condition is very non-trivial evidence for the internal consistency of string/M-theory. 

In the analysis leading to the M9 brane, supersymmetry plays a central role. Recently, there has been a renewed interest in non-supersymmetric string theories in ten dimensions (see e.g. \cite{BoyleSmith:2023xkd, Basile:2023knk,Abel:2024vov,Cribiori:2021txm,Raucci:2022bjw,Baykara:2022cwj,Koga:2022qch,Angelantonj:2023egh,Nakajima:2023zsh,Mourad:2023wjg,Matyas:2023nds,Avalos:2023ldc,Fraiman:2023cpa,Mourad:2023loc,DeFreitas:2024ztt,Tachikawa:2024ucm,Saxena:2024eil,Baykara:2024tjr,Mourad:2024dur,Mourad:2024mpg,Angelantonj:2024jtu,Detraux:2024esd,Leone:2024xae,Raucci:2024fnp,Larotonda:2024thv,Basile:2025lek}), but there has been little progress in similar questions in M-theory, due to its non-perturbative nature. In this note, we will take a small step in this direction by asking and partially answering the question of whether, if we drop supersymmetry, there may exist additional, \emph{exotic} end-of-the-world branes in M-theory. 

Due to anomaly inflow constraints, any exotic end-of-the-world branes that may exist must still have a massless, chiral spectrum, although this may be non-supersymmetric, and this provides us with a handle to tackle the problem even in absence of a perturbative description. Specifically, we will work out the general consistency conditions imposed in the chiral spectrum by anomaly inflow, but not require that the boundary gauge theory is supersymmetric, and search for non-trivial solutions. 

One might have thought that the answer to this question is known indirectly, since exotic end-of-the-world branes for M theory are intrinsically connected to non-supersymmetric heterotic strings in ten dimensions. Specifically, for any given end-of-the-world brane in M-theory,  one could  take two of them and form an interval compactification\footnote{We note in passing that the $SO(16)^2$ string \cite{Alvarez-Gaume:1986ghj, Dixon:1986iz} cannot possibly become an interval compactification of M-theory at strong coupling in this way, since its anomaly polynomial does not have a form compatible with M-theory anomaly inflow.} (as one does in the supersymmetric context \cite{Horava:1995qa,Horava:1996ma}). Then, upon shrinking the interval size, this would yield perturbative heterotic strings, which are fully classified \cite{BoyleSmith:2023xkd}. The problem with this argument is that, since the exotic end of the world branes would be non-supersymmetric, one should expect them to source a tension, in which case there could be a potential obstructing the limit of zero interval size. The relationship between interval size and coupling could be altered due to warping, leading to a heterotic-like string, but non-perturbative. Such strings may exist or not, but if they do, they cannot be captured by a perturbative worldsheet classification. Therefore, the question we ask here is quite general, and the exotic end-of-the-world branes do not have to be directly related in a direct way to heterotic strings.

As described in the rest of the note, we solve the problem of anomaly inflow exhaustively for a large class of simple gauge groups, under the simplifying assumption of a purely chiral spectrum, and find no new boundary conditions for M-theory. We also perform a large search of non-supersymmetric boundary conditions for the exceptional groups, and find a novel solution with gauge group $G_2$ and just three non-trivial matter representations. Anomaly inflow leads to five conditions on the anomaly polynomial in this case, so the existence of a solution involving only three charged matter fields is somewhat nontrivial. We do not know whether it is of any physical significance, but find its existence interesting. We also uncovered two more solutions, with fully chiral spectrum and gauge groups $G_2$ and $Sp(2)$ respectively, where the anomaly polynomial factorizes (so that anomalies can be cancelled) but not in a way compatible with M-theory; the solutions might perhaps be relevant for other corners of the string Landscape.

The remainder of this paper has the following structure. In Section \ref{sec:review} we examine the anomaly inflow of the topological M-theory terms on a single boundary, reviewing the well-known supersymmetric $E_8$ M9-brane. In Section \ref{sec:searchbc} we neglect supersymmetry and carry out a similar analysis for a $G$ gauge theory, with $G$ a simple Lie group, obtaining a set of linear equations that allow us to determine the appropriate boundary matter content for anomaly cancellation. From this, we give explicit results for various Lie algebras. Finally, we conclude in Section \ref{sec:conclusion} and briefly comment on possible future directions.

There are two appendices. In Appendix \ref{appA} we summarize a set of anomaly polynomials for $G_2$ and $Sp(6)$, with appealing Green-Schwarz factorization, which however cannot be emebedded within M-theory.  Appendix \ref{App:B} contains technical details on computing indices of representations of relevance to our calculations.

\section{Review of anomaly inflow on M9 brane}\label{sec:review}
In this Section we review the miraculous anomaly inflow mechanism in the Ho\v{r}ava-Witten compactification of M-theory, \cite{Horava:1995qa, Horava:1996ma}. In these references, the compactification of M-theory on an interval $S^1/\mathbb{Z}_2$ (where the $\mathbb{Z}_2$ acts by reflecting the $S^1$ coordinate) is discussed. Compactification of M-theory on a manifold in boundary, such as an interval, is potentially problematic for the following reason. At low energies, M-theory is well approximated by 11d supergravity coupled together with a topological higher-derivative term,  
\begin{equation}
\label{eq:topcouMtheory}
     S \supset 2 \pi \mathsf{i} \int_{Y_{11}} \left(\frac{1}{6}\,C_3 \wedge G_4 \wedge G_4 - C_3\wedge I_8  \right),
\end{equation}
where $I_8$ is a particular characteristic polynomial in terms of the first and second Pontryagin classes. Its explicit form is 
\begin{equation}
\label{eq:8form}
I_8 = \frac{1}{48}\left[p_2-\frac{1}{4} p_1^2 \right].
\end{equation}
The quantities $p_i$ (the Pontryagin classes) are certain combinations of powers of traces of the Riemann tensor; our conventions for these, as well as a quick survey of our techniques to study anomaly cancellation, are given in Appendix \ref{App:B}.
A priori, it is not clear that \eqref{eq:topcouMtheory} is well-defined, since it involves fractional coefficients for the Chern-Simons like couplings. Whenever $Y_{11}$ together with the  three-form profile $C_3$ can be extended to a 12-dimensional manifold $Z_{12}$, such that $Y_{11}=\partial Z_{12}$, one can use Stokes' theorem to write the term \eqref{eq:topcouMtheory} as
\begin{equation}
\label{eq:topcouMtheory3}
      2 \pi \mathsf{i} \int_{Z_{12}} \left(\frac{1}{6}\,G_4 \wedge G_4 \wedge G_4 - G_4\wedge I_8  \right),
\end{equation}
where the integral is evaluated on a manifold with boundary. The problem is now to show that this expression is independent of the choice of $Z_{12}$ and extension of the three-form field. To do this, one needs to show that \eqref{eq:topcouMtheory} is properly quantized on closed 12-dimensional manifolds, a feat achieved in full generality in \cite{Witten:1996md}, and only in combination with a gravitino contribution.

In any case, the term is still problematic on a manifold with boundary, such as an interval;  \eqref{eq:topcouMtheory} is not invariant under gauge transformations of the M-theory three form $C_3$; one picks an anomalous variation of the action, 
\begin{equation}
\label{eq:topcouMtheory2}
     \delta_{\Lambda_2}S \supset 2 \pi \mathsf{i} \int_{\partial Y_{11}} \Lambda_2 \wedge\left(\frac{1}{6}\,G_4 \wedge G_4 - I_8  \right),
\end{equation}
where $\Lambda_2$ is the gauge parameter. 

The deep insight of \cite{Horava:1995qa} was that this anomalous variation may be cancelled by the addition of suitable chiral degrees of freedom at the M-theory boundaries. Specifically, reference \cite{Horava:1995qa} considers ten-dimensional chiral fermions charged under a gauge group $G$. Their perturbative anomaly is encoded in terms of a twelve-dimensional anomaly polynomial
\begin{equation} \int_{Z_{12}} \mathcal{P}(R,F_G),\label{angauge}\end{equation}
which we may add to \eqref{eq:topcouMtheory3} to obtain the total anomaly theory of M-theory on a manifold with boundary. At first sight, \eqref{angauge} and \eqref{eq:topcouMtheory3} are very different -- they even involve different gauge potentials. However, one may impose a boundary condition
\begin{equation}
 G_4\vert_{\partial Y_{11}}=a\, c_2 + b\, p_1,\label{bc}
\end{equation}
where $a,b$ are so far undetermined coefficients; $p_1$ is the first Pontryagin class of the tangent bundle of the boundary, and $c_2$ is the second Chern class of the $G$ gauge fields living there in some reference representation. The identification is supposed to hold at the level of cohomology only, but substituting $G_4$ by its expression in terms of gauge and gravitational characteristic classes, the anomalous theory \eqref{eq:topcouMtheory3} takes the form 
 \begin{equation}
\label{eq:topcouMtheory4}
      2 \pi \mathsf{i} \int_{Z_{12}} \left[\frac{1}{6}\left(a\, c_2 + b\, p_1\right)^2  -  I_8  \right]\wedge \left(a\, p_1 + b\, c_2\right).
\end{equation}
This looks exactly like a gauge-gravity anomaly polynomial -- albeit a very special one, where only the class $c_2$ (and no higher Chern classes) appear, and which factorizes in a precise way. 

In \cite{Horava:1995qa, Horava:1996ma}, it was further assumed that the boundary theory preserves $\mathcal{N}=1$ supersymmetry. This means that the only chiral fermions that can appear are gaugini, so the whole boundary theory is fixed by the choice of gauge group $G$. Remarkably, for two copies $G=E_8$ (corresponding to the two endpoints of the interval),
\begin{equation}
    \label{Eq:OneTotAn}
 \frac12 \mathcal{P}_{\mathrm{Grav}}(R)+\mathcal{P}_{E_8}(R,F) = -\frac{1}{96}\left( \frac{c_{2}}{15} + p_1\right)\left[\frac{1}{4}\left( \frac{c_{2}}{15} + p_1\right)^2 + \frac{1}{8} p_1^2 - \frac{1}{2} p_2\right],
 \end{equation}
which is \emph{exactly} of the form \eqref{eq:topcouMtheory4} for $a=1/60, b=1/4$ (where $c_2$ is the Chern class in the adjoint representation of $E_8$) once \eqref{eq:8form} is used, thereby cancelling perturbative anomalies. The term $\frac{1}{2} \mathcal{P}_{\mathrm{Grav}}(R)$ corresponds to the contribution of the 10d chiral fields coming from the 11d gravitino reduced on the interval. 

In short, supersymmetry and anomaly cancellation force the boundary of an M-theory compactification to have $E_8$ gauge fields, and there is no other option. The point of this paper is to find out whether there are more possibilities if the requirement of supersymmetry is dropped. This analysis is carried out in detail in Section \ref{sec:searchbc}. We will end this Section by describing this construction in modern language: an anomalous field theory in $d$ dimensions is associated to an invertible field theory in $(d+1)$ dimensions \cite{Freed:2014iua, Freed:2016rqq}, the anomaly theory \cite{Monnier:2019ytc}; the anomalous theory is realized as a boundary condition for the anomaly theory. In this case, the chiral degrees of freedom live in ten dimensions and the anomaly theory is, using the identification \eqref{bc}, the M-theory topological couplings \cite{Witten:1996md}. Although these terms are topological in the sense that they do not depend on the metric explicitly, they do depend continuously on the three-form profile, as befits the anomaly theory for a perturbative symmetry. From this point of view, the $E_8$ gauge fields are simply the suitable boundary degrees of freedom on a boundary required by the bulk topological couplings of M-theory. 

\section{Searching for more general boundary conditions} \label{sec:searchbc}
In the previous Section, we described how the combined requirements of anomaly inflow and supersymmetry uniquely fix the matter content of a boundary in M-theory to be $E_8$ gauge fields. The anomaly polynomial of the gauge fields exactly matches the M-theory topological couplings \eqref{eq:topcouMtheory}. 

The main question we wish to address in this paper is how this picture is modified, if at all, when the requirement of boundary supersymmetry is dropped. The role of supersymmetry is fixing the matter content uniquely (to a single copy of the adjoint representation) as a function of the gauge group $G$ of the boundary. Without it, and still focusing on the case of a simple Lie group $G$, we must consider chiral fermions transforming in a general matter representation
\begin{equation}
\label{eq:combiofreps}
\underbrace{(\mathsf{r}_1) \oplus \cdots \oplus (\mathsf{r}_1)}_{n_{\mathsf{r}_1}\rm\ times} \oplus \underbrace{(\mathsf{r}_2) \oplus \cdots \oplus (\mathsf{r}_2)}_{n_{\mathsf{r}_2}\rm\ times} \oplus \underbrace{(\mathsf{r}_3) \oplus \cdots \oplus (\mathsf{r}_3)}_{n_{\mathsf{r}_3}\rm\ times} \oplus \cdots
\end{equation}
where~\eqref{eq:combiofreps} means that we will take $n_{\mathsf{r}_1}$ multiplets of the (irreducible) $\mathsf{r}_1$ representation, $n_{\mathsf{r}_2}$ multiplets of the $\mathsf{r}_2$ representation, $n_{\mathsf{r}_3}$ multiplets of the $\mathsf{r}_3$ and so forth. The corresponding 12d anomaly polynomial is
\begin{align}
\begin{split}
\frac{1}{2} \mathcal{P}_{\mathrm{Grav}}(R) + \sum_{\mathsf{r}_i} \mathcal{P}_{G}(R,F_{\mathsf{r}_i}) & = \frac{1}{2} \sum_{\mathsf{r}_i} \mathrm{ch}_{6,\mathsf{r}_i} - \frac{1}{48}\, p_1\, \sum_{\mathsf{r}_i} \mathrm{ch}_{4,\mathsf{r}_i} + \frac{1}{11520} (7\, p_1^2 - 4\, p_2)\,\sum_{\mathsf{r}_i} \mathrm{ch}_{2,\mathsf{r}_i} \\ 
& + \frac{1}{2}\frac{1}{967680}\left[\left( 128-31\,\sum_{\mathsf{r}_i} \mathrm{ch}_{0,\mathsf{r}_i} \right)\,p_1^3 \right.\\  
& \qquad \qquad \qquad + \left.\left(44\,\sum_{\mathsf{r}_i} \mathrm{ch}_{0,\mathsf{r}_i} -832 \right)\,p_1\,p_2 \right. \\ 
& \qquad \qquad \qquad \qquad \quad + \left. \left(3968 - 16\,\sum_{\mathsf{r}_i} \mathrm{ch}_{0,\mathsf{r}_i} \right)\, p_3 \right]\,,
\end{split}
\label{indexsg}\end{align}
where $\mathrm{ch}_{l,\mathsf{r}_i}$ represents the $2l$-form piece of the Chern character of representation $\mathsf{r}_i$ (see Appendix \ref{App:B}). Notice that several terms are absent due to degree reasons; for instance, there is no term with $l$ odd, as they cannot be combined with a gravitational class to produce a term of degree 12.

The basic question we wish to answer is under which conditions does the anomaly polynomial \eqref{indexsg} factorize in the form \eqref{eq:topcouMtheory4} suitable for M-theory anomaly cancellation. Clearly, several coefficients must vanish. For instance, from~\eqref{eq:combiofreps} and properties of the Chern character, it follows that $\sum_{\mathsf{r}_i} \mathrm{ch}_{0,\mathsf{r}_i} = \sum_{\mathsf{r}_i} n_{\mathsf{r}_i}\,\text{dim}(\mathsf{r}_i)$. Therefore, the last term on the right-hand side of equation~\eqref{indexsg} certainly vanishes as long as
\begin{equation} \label{eq:dim248}
\sum_ i n_{\mathsf{r}_i} \text{dim}(\mathsf{r}_i)=248,
\end{equation}
i.e. the total number of fermions, counted according to their chirality, should be $248$.

To study the implications that factorization has for our problem, we will introduce some notation. In general, Chern characters in different representations are not identical, but they can be related to each other through group-theoretic factors \cite{Okubo:1981td, 10.1063:1.525670}. Thus, very generally, one can write
\begin{equation} \label{chernexpan}\mathrm{ch}_{l,\mathsf{r}_i}=\sum_{k}\mathrm{u}_{\mathsf{r}_i}^{lk} \mathrm{ch}_{l-k}\,  \mathrm{ch}_{k},\end{equation}
where $\mathrm{ch}_{k}$ denotes the components of the Chern character in some reference representation, and $\mathrm{u}_{\mathsf{r}_i}^{lk}$ is a matrix of coefficients. Notice that, unlike in the anomaly polynomial \eqref{indexsg}, these expressions generally involve both even and odd components of the Chern character in the reference representation. These matrix coefficients are related to a set of group-theoretical constants related to the Dynkin index or equivalently to Casimir invariants of representations described and computed in Appendix \ref{App:B}. Therefore, using our general arrangement of matter \eqref{eq:combiofreps}, properties of the Chern character such as $\sum_{\mathsf{r}_i} \mathrm{ch}_{l,\mathsf{r}_i} = \sum_{\mathsf{r}_i} n_{\mathsf{r}_i} \ch_{l, \mathsf{r}}$, equations \eqref{eq:traces}, for an arbitrary representation $\mathsf{r}$, we can express $\ch_{l, \mathsf{r}}$ as combinations of Chern characters of the same and less order in some reference representation via \eqref{chernexpan}. For instance, for the sixth-order character, we obtain 
\begin{align}
\begin{split}
    \ch_{6,\mathsf{r}} & = \mathrm{u}_{\mathsf{r}}^{(1)}\, \ch_6 + \mathrm{u}_{\mathsf{r}}^{(2)}\, \ch_5 \,\ch_1 + \mathrm{u}_{\mathsf{r}}^{(3)}\,  \ch_4 \, \ch_2 + \mathrm{u}_{\mathsf{r}}^{(4)}\, \ch_4 \,\ch_1^2 + \mathrm{u}_{\mathsf{r}}^{(5)}\, \ch_3^2 + \mathrm{u}_{\mathsf{r}}^{(6)}\, \ch_3\,\ch_2\,\ch_1  \\
    & \hspace{1.5cm} + \mathrm{u}_{\mathsf{r}}^{(7)}\, \ch_3\,\ch_1^3 + \mathrm{u}_{\mathsf{r}}^{(8)}\, \ch_2^3 + \mathrm{u}_{\mathsf{r}}^{(9)}\, \ch_2^2\,\ch_1^2 + \mathrm{u}_{\mathsf{r}}^{(10)}\, \ch_2\,\ch_1^4 + \mathrm{u}_{\mathsf{r}}^{(11)}\, \ch_1^6 \,, 
\end{split}
\end{align}
where we have altered our notation to make it slightly less cumbersome, relabeling $\mathrm{u}_{\mathsf{r}}^{0,6}$ as $\mathrm{u}_{\mathsf{r}}^{(1)}$, $\mathrm{u}_{\mathsf{r}}^{1,5}$ as $\mathrm{u}_{\mathsf{r}}^{(2)}$, and so forth. Similarly, we have expressions for the fourth- and second-order characters, which can be expressed as
\begin{align}
    \ch_{4,\mathsf{r}} & = \mathrm{u}_{\mathsf{r}}^{(12)}\, \ch_4 + \mathrm{u}_{\mathsf{r}}^{(13)}\, \ch_3\,\ch_1 + \mathrm{u}_{\mathsf{r}}^{(14)}\, \ch_2^2 + \mathrm{u}_{\mathsf{r}}^{(15)}\, \ch_2\,\ch_1^2 + \mathrm{u}_{\mathsf{r}}^{(16)}\, \ch_1^4 \,, \\
    \ch_{2,\mathsf{r}} & = \mathrm{u}_{\mathsf{r}}^{(17)}\, \ch_2 + \mathrm{u}_{\mathsf{r}}^{(18)}\, \ch_1^2.
\end{align}
The vectors $\mathrm{u}_{\mathsf{r}}^{(i)}, i \in \{ 1, \dots , 18\}$ are the nonzero group-theory coefficients relevant to our problem, and they are a function only of the representation $\mathsf{r}$.
Depending on the algebra under consideration, some of these coefficients may vanish identically. For instance, for simple Lie algebras, we know that $\mathrm{tr}_r (F) = 0$, whereas, for real representations $\mathrm{tr}_r (F^l)$ vanishes for $l$ odd.

Using these expressions, the anomaly polynomial \eqref{indexsg} can be rewritten as
\begin{align}
\begin{split}
\mathcal{P}_G(R,F) & = \frac{1}{2} \sum_{\mathsf{r}_i} n_{\mathsf{r}_i} \left( \mathrm{u}_{\mathsf{r}_i}^{(1)}\, \ch_6 + \mathrm{u}_{\mathsf{r}_i}^{(2)}\, \ch_5 \,\ch_1 + \mathrm{u}_{\mathsf{r}_i}^{(3)}\,  \ch_4 \, \ch_2 + \mathrm{u}_{\mathsf{r}_i}^{(4)}\, \ch_4 \,\ch_1^2 + \mathrm{u}_{\mathsf{r}_i}^{(5)}\, \ch_3^2  + \right. \\ & \left. \mathrm{u}_{\mathsf{r}_i}^{(6)}\, \ch_3\,\ch_2\,\ch_1   
+ \mathrm{u}_{\mathsf{r}_i}^{(7)}\, \ch_3\,\ch_1^3 + \mathrm{u}_{\mathsf{r}_i}^{(8)}\, \ch_2^3 + \mathrm{u}_{\mathsf{r}_i}^{(9)}\, \ch_2^2\,\ch_1^2 + \mathrm{u}_{\mathsf{r}_i}^{(10)}\, \ch_2\,\ch_1^4 + \mathrm{u}_{\mathsf{r}_i}^{(11)}\, \ch_1^6 \right) \\[5pt]
& - \frac{1}{48}\, p_1 \sum_{\mathsf{r}_i} n_{\mathsf{r}_i} \left(\mathrm{u}_{\mathsf{r}_i}^{(12)}\, \ch_4 + \mathrm{u}_{\mathsf{r}_i}^{(13)}\, \ch_3\,\ch_1 + \mathrm{u}_{\mathsf{r}_i}^{(14)}\, \ch_2^2 + \mathrm{u}_{\mathsf{r}_i}^{(15)}\, \ch_2\,\ch_1^2 + \mathrm{u}_{\mathsf{r}_i}^{(16)}\, \ch_1^4 \right)  
\\ & + \frac{1}{11520} (7 p_1^2 - 4 p_2) \sum_{\mathsf{r}_i} n_{\mathsf{r}_i} \left( \mathrm{u}_{\mathsf{r}_i}^{(17)}\, \ch_2 + \mathrm{u}_{\mathsf{r}_i}^{(18)}\ch_1^2\right) + \frac{1}{192} p_1 p_2 - \frac{1}{256} p_1^3\,,
\end{split}\label{Eq:GGenAnPol}
\end{align}
where we have denoted $\frac{1}{2} \mathcal{P}_{\mathrm{Grav}}(R) + \sum_{\mathsf{r}_i} \mathcal{P}_{G}(R,F_{\mathsf{r}_i})$ as $\mathcal{P}_G(R,F)$ for brevity. Clearly, many terms have to be cancelled to achieve the factorized form of Section \ref{sec:review}, which is
\begin{equation}\label{eq:factAPG}
    \mathcal{P}_G(R,F) = -\frac{1}{96}(-a\, \ch_2 + p_1)\left[ \frac{1}{4}(-a\, \ch_2 + p_1)^2 + \frac{1}{8} p_1^2 - \frac{1}{2} p_2 \right].
\end{equation}
Here, we have set the coefficient $b=1$, since it is fixed to that value by demanding matching of the pure gravitational anomaly (specifically, the $p_1^3$ or $p_1\, p_2$ terms). As a result, the boundary condition \eqref{bc} depends only on $a$.
Imposing \eqref{eq:factAPG} also leads to a system of linear conditions, which we present into two groups. The first one is
\begin{align}\label{ffr}
\begin{split}
     n_{\mathsf{r}_1} \mathrm{u}_{\mathsf{r}_1}^{(2)} + n_{\mathsf{r}_2} \mathrm{u}_{\mathsf{r}_2}^{(2)} + n_{\mathsf{r}_3}\mathrm{u}_{\mathsf{r}_3}^{(2)} + \cdots & = 0 \\
     n_{\mathsf{r}_1} \mathrm{u}_{\mathsf{r}_1}^{(4)} + n_{\mathsf{r}_2} \mathrm{u}_{\mathsf{r}_2}^{(4)} + n_{\mathsf{r}_3}\mathrm{u}_{\mathsf{r}_3}^{(4)} +  \cdots & = 0 \\
     n_{\mathsf{r}_1} \mathrm{u}_{\mathsf{r}_1}^{(6)} + n_{\mathsf{r}_2} \mathrm{u}_{\mathsf{r}_2}^{(6)} + n_{\mathsf{r}_3}\mathrm{u}_{\mathsf{r}_3}^{(6)} + \cdots & = 0 \\
     n_{\mathsf{r}_1} \mathrm{u}_{\mathsf{r}_1}^{(7)} + n_{\mathsf{r}_2} \mathrm{u}_{\mathsf{r}_2}^{(7)} + n_{\mathsf{r}_3}\mathrm{u}_{\mathsf{r}_3}^{(7)} +  \cdots & = 0 \\
     n_{\mathsf{r}_1} \mathrm{u}_{\mathsf{r}_1}^{(9)} + n_{\mathsf{r}_2} \mathrm{u}_{\mathsf{r}_2}^{(9)} + n_{\mathsf{r}_3}\mathrm{u}_{\mathsf{r}_3}^{(9)} + \cdots & = 0 \\
     n_{\mathsf{r}_1} \mathrm{u}_{\mathsf{r}_1}^{(10)} + n_{\mathsf{r}_2} \mathrm{u}_{\mathsf{r}_2}^{(10)} + n_{\mathsf{r}_3}\mathrm{u}_{\mathsf{r}_3}^{(10)} +  \cdots & = 0 \\
     n_{\mathsf{r}_1} \mathrm{u}_{\mathsf{r}_1}^{(11)} + n_{\mathsf{r}_2} \mathrm{u}_{\mathsf{r}_2}^{(11)} + n_{\mathsf{r}_3}\mathrm{u}_{\mathsf{r}_3}^{(11)} + \cdots & = 0 \\
     n_{\mathsf{r}_1} \mathrm{u}_{\mathsf{r}_1}^{(13)} + n_{\mathsf{r}_2} \mathrm{u}_{\mathsf{r}_2}^{(13)} + n_{\mathsf{r}_3}\mathrm{u}_{\mathsf{r}_3}^{(13)} +  \cdots & = 0 \\
     n_{\mathsf{r}_1} \mathrm{u}_{\mathsf{r}_1}^{(15)} + n_{\mathsf{r}_2} \mathrm{u}_{\mathsf{r}_2}^{(15)} + n_{\mathsf{r}_3}\mathrm{u}_{\mathsf{r}_3}^{(15)} + \cdots & = 0 \\
     n_{\mathsf{r}_1} \mathrm{u}_{\mathsf{r}_1}^{(16)} + n_{\mathsf{r}_2} \mathrm{u}_{\mathsf{r}_2}^{(16)} + n_{\mathsf{r}_3}\mathrm{u}_{\mathsf{r}_3}^{(16)} +  \cdots & = 0 \\
     n_{\mathsf{r}_1} \mathrm{u}_{\mathsf{r}_1}^{(18)} + n_{\mathsf{r}_2} \mathrm{u}_{\mathsf{r}_2}^{(18)} + n_{\mathsf{r}_3}\mathrm{u}_{\mathsf{r}_3}^{(18)} + \cdots & = 0\,.
\end{split}
\end{align}
and it comes from demanding matching of all the terms with a $\text{ch}_1$ piece. As we remarked above, for simple groups $\text{ch}_1=0$, and so in this case the conditions \eqref{ffr} will be satisfied automatically. Relatedly, since there is no $\text{ch}_1$ in any representation, all coefficients $\mathrm{u}_{\mathsf{r}}^{(i)}$ involving $\text{ch}_1$ will vanish identically. These correspond to 
\begin{equation}i=2,4,6,7,9,11,13,15,16,18,\end{equation}
which are precisely the variables appearing in \eqref{ffr}.
Conversely, the second group consists of conditions without a factor of $\ch_1$, and that therefore apply to simple algebras as well. This consists of four homogeneous conditions:
\begin{align}\label{eq:lsforsimplealg}
\begin{split}
n_{\mathsf{r}_1} \mathrm{u}_{\mathsf{r}_1}^{(1)} + n_{\mathsf{r}_2} \mathrm{u}_{\mathsf{r}_2}^{(1)} + n_{\mathsf{r}_3}\mathrm{u}_{\mathsf{r}_3}^{(1)} +  \cdots & = 0 \\
n_{\mathsf{r}_1} \mathrm{u}_{\mathsf{r}_1}^{(3)} + n_{\mathsf{r}_2} \mathrm{u}_{\mathsf{r}_2}^{(3)} + n_{\mathsf{r}_3}\mathrm{u}_{\mathsf{r}_3}^{(3)} + \cdots & = 0 \\
n_{\mathsf{r}_1} \mathrm{u}_{\mathsf{r}_1}^{(5)} + n_{\mathsf{r}_2} \mathrm{u}_{\mathsf{r}_2}^{(5)} + n_{\mathsf{r}_3}\mathrm{u}_{\mathsf{r}_3}^{(5)} + \cdots & = 0 \\
n_{\mathsf{r}_1} \mathrm{u}_{\mathsf{r}_1}^{(12)} + n_{\mathsf{r}_2} \mathrm{u}_{\mathsf{r}_2}^{(12)} + n_{\mathsf{r}_3}\mathrm{u}_{\mathsf{r}_3}^{(12)} + \cdots & = 0 \,.
\end{split}
\end{align}
as well as four inhomogeneous equations,
\begin{align}\label{eq:fermcounting}\begin{split}
n_{\mathsf{r}_1}\mathrm{u}_{\mathsf{r}_1}^{(0)} + n_{\mathsf{r}_2}\mathrm{u}_{\mathsf{r}_1}^{(0)} + n_{\mathsf{r}_3}\mathrm{u}_{\mathsf{r}_3}^{(0)} + \cdots & = 248\\
   n_{\mathsf{r}_1}\, \mathrm{u}_{\mathsf{r}_1}^{(8)} + n_{\mathsf{r}_2}\, \mathrm{u}_{\mathsf{r}_2}^{(8)} + n_{\mathsf{r}_3}\, \mathrm{u}_{\mathsf{r}_3}^{(8)} + \cdots & = \frac{15}{32} a^3 \\
     n_{\mathsf{r}_1} \mathrm{u}_{\mathsf{r}_1}^{(14)} + n_{\mathsf{r}_2} \mathrm{u}_{\mathsf{r}_2}^{(14)} + n_{\mathsf{r}_3} \mathrm{u}_{\mathsf{r}_3}^{(14)} + \cdots & = \frac{9}{4} a^2 \\
     n_{\mathsf{r}_1} \mathrm{u}_{\mathsf{r}_1}^{(17)} + n_{\mathsf{r}_2} \mathrm{u}_{\mathsf{r}_2}^{(17)} + n_{\mathsf{r}_3} \mathrm{u}_{\mathsf{r}_3}^{(17)} +\cdots & = 15 a,
\end{split}
\end{align}
where we have denoted $\mathrm{dim}(\mathsf{r}_i) = \mathrm{u}_{\mathsf{r}_i}^{(0)}$. Note that the first equation of \eq{eq:fermcounting} is merely the restriction that the total fermion number, counted by chiralities, is $248$.

As mentioned above, if we restrict ourselves to simple algebras the system of linear conditions is reduced to \eqref{eq:lsforsimplealg} and \eqref{eq:fermcounting}, and the equations in \eqref{ffr} can be ignored; we will restrict to this case for the remainder of this draft, for simplicity, and only consider boundary conditions for M-theory where the boundary gauge algebra is simple. In fact, even in this case, we will find further vanishing coefficients depending on the explicit algebra under consideration. For example, for the exceptional algebras there are no genuine fourth-order Casimir invariants \cite{Okubo:1981td} as well as odd-order Casimir (except for $E_6$ that has fifth-order and ninth-order Casimirs), thus $\mathsf{u}_{\mathrm{r}_i}^{(3)}$, $\mathsf{u}_{\mathrm{r}_i}^{(5)}$, $\mathsf{u}_{\mathrm{r}_i}^{(12)}$ vanish. This simplifies the system \eqref{eq:lsforsimplealg} even more. 
On the other hand, for the simple algebras $B_n$, $C_n$ and $D_n$  there are no odd-order Casimir invariants, so $\mathsf{u}_{\mathrm{r}_i}^{(3)}$ is always zero.

In the general simple case, without further vanishings, we therefore have a system of eight equations, in as many variables $n_{\mathsf{r}_1}$ as there are representations included, plus the coefficient $a$. For more than eight representations, there will be infinitely many solutions, and therefore the system is underdetermined. Furthermore, to any given solution we can add any non-chiral (and therefore, non-anomalous) spectrum that we wish, and the result will also solve the system of equations. 

This large ambiguity in the space of solutions is unlikely to be physically meaningful, and we need some way to cut it down. In this note, we will do this by additionally demanding that all chiral fields must have the same chirality, and that the gauge algebra $G$ contains a single simple factor, as is the case for the Ho\v{r}ava-Witten $E_8$ solution. While this is a significant constraint in the space of solutions (and, without supersymmetry, there is no reason why we could not involve fields with different chiralities), it has the advantage that since the dimensions of the representations have to add up to $248$, the space of potential solutions is bounded, and as we will see, can be explored fully for some groups. We remark that, if the restriction of chirality is dropped, then there are additional solutions (we will give one such example for $G_2$ below), but we have not explored this larger space in detail.

\subsection{\texorpdfstring{$G_2$}{G2} gauge group}\label{sec:G2}

We will first ask the question for the exceptional group $G_2$. This group only has real representations, several of which have dimension less than 248. Thus, one might expect several different combinations of representations that solve the linear system of equations in~\eqref{eq:lsforsimplealg} and \eqref{eq:fermcounting}. To explore this, we apply the algorithm described previously.  

Being a real, semisimple algebra, all odd powers of the Chern character vanish. Furthermore, for exceptional algebras such as $G_2$ there is no modified fourth-order index \cite{Okubo:1981td}, which means that a fourth-order Casimir is proportional to powers of a lower-order Casimir. Thus, the general system \eqref{eq:lsforsimplealg} simplifies significantly. We need the sixth-order indices of these representations found in~\cite{vanRitbergen:1998pn} to proceed. These are related to the constants $\mathrm{u}_{\mathsf{r}}^{(1)}$ showing up in the trace identity $\mathrm{tr}_{\mathsf{r}}\,F^6  = {\mathrm{u}}_{\mathsf{r}}^{(1)}\mathrm{tr}\,F^6 + {\mathrm{u}}_{\mathsf{r}}^{(8)}(\mathrm{tr}\,F^2)^3$ (up to an irrelevant normalization of $1/6!$) needed to re-express the sixth-order piece of the Chern character in an arbitrary representation in terms of the Chern character of the fundamental representation of $G_2$, which we choose as our reference representation (see the discussion of Section \ref{sec:searchbc}). The second-order indices needed to determine the other constants appearing in~\eqref{Eq:GGenAnPol}, have also been computed in many places (see e.g. \cite{Yamatsu:2015npn}). 
We note in passing that these specific trace identities are only valid for~\cite{Okubo:1981td} $A_1, A_2$ and exceptional algebras. Taking all of this into account, we find that only five non-trival anomaly cancellation constraints remain.

Following these steps, we have identified all possible solutions to this system where the degeneracies are positive, i.e. every particle has positive chirality. There are only four possibilities, shown in table~\ref{table1} which depicts the spectrum and the corresponding factorized anomaly polynomial. It turns out that all of these solutions arise from restriction of a single $E_8$ adjoint to different branches of $G_2$ subgroups; as a result, they do not constitute qualitatively new boundary conditions for M-theory, but rather correspond merely to taking the Ho\v{r}ava-Witten M9 and looking at anomalies in a $G_2$ subgroup of $E_8$. The first spectrum comes from the branching \cite{Yamatsu:2015npn} \begin{equation}E_8 \supset G_2 \times F_4,\quad  \mathsf{248} \rightarrow (\mathsf{14},\mathsf{1})\oplus (\mathsf{7},\mathsf{26}) \oplus (\mathsf{1},\mathsf{52}).\end{equation} 
The second solution can be similarly found by the chain of embeddings
\begin{equation} E_8 \supset E_7 \otimes SU(2),\quad E_7 \supset SU(2) \times G_2,\end{equation}
and the third corresponds to 
\begin{equation} E_8 \supset E_6 \otimes SU(3),\quad E_6 \supset G_2.\end{equation}
Finally, the fourth comes from the chain
\begin{equation} E_8\supset\, SO(14)\supset\, G_2.\end{equation}

\begin{table}[!ht]
\centering
\renewcommand{\arraystretch}{2.5}
\begin{tabular}{c|c|c}
\hline
\multicolumn{3}{c}{$G_2$ gauge group} \\ \hline
& $(n_{\mathsf{1}}=52, n_{\mathsf{7}}=26, n_{\mathsf{14}}=1)$ & \\
$\mathcal{P}_{G_2}(R,F)=$ & $\displaystyle{-\frac{1}{96}(2\,c_{2,\mathsf{7}} + p_1)\left(\frac{1}{4}(2\,c_{2,\mathsf{7}} + p_1)^2 + \frac{1}{8}\,p_1^2 - \frac{1}{2}\,p_2 \right)}$ & $\displaystyle{G_4 = \frac{1}{4}\,p_1 + \frac{1}{2}\,c_{2,{\mathsf{7}}}}$ \\ \hline 
& $(n_{\mathsf{1}}=6, n_{\mathsf{7}}=13, n_{\mathsf{14}}=5, n_{\mathsf{27}}=3)$ & \\
$\mathcal{P}_{G_2}(R,F)=$ & $\displaystyle{-\frac{1}{96}(4\,c_{2,\mathsf{7}} + p_1)\left(\frac{1}{4}(4\,c_{2,\mathsf{7}} + p_1)^2 + \frac{1}{8}\,p_1^2 - \frac{1}{2}\,p_2 \right)}$ & $\displaystyle{G_4 = \frac{1}{4}\,p_1 + c_{2,{\mathsf{7}}}}$ \\ \hline 
&  $(n_{\mathsf{1}}=8, n_{\mathsf{14}}=1, n_{\mathsf{27}}=6, n_{\mathsf{64}}=1)$ & \\
$\mathcal{P}_{G_2}(R,F)=$ & $\displaystyle{ -\frac{1}{96}(6\,c_{2,\mathsf{7}} + p_1)\left(\frac{1}{4}(6\,c_{2,\mathsf{7}} + p_1)^2 + \frac{1}{8}\,p_1^2 - \frac{1}{2}\,p_2 \right)}$ & $\displaystyle{G_4 = \frac{1}{4}\,p_1 + \frac{3}{2}\,c_{2,{\mathsf{7}}}}$ \\ \hline 
&  $(n_{\mathsf{1}}=1, n_{\mathsf{14}}=3, n_{\mathsf{64}}=2, n_{\mathsf{77}}=1)$ & \\ 
$\mathcal{P}_{G_2}(R,F)=$ & $ \displaystyle{ -\frac{1}{96}(8\,c_{2,\mathsf{7}} + p_1)\left(\frac{1}{4}(8\,c_{2,\mathsf{7}} + p_1)^2 + \frac{1}{8}\,p_1^2 - \frac{1}{2}\,p_2 \right)}$ & $\displaystyle{G_4 = \frac{1}{4}\,p_1 +  2\,c_{2,{\mathsf{7}}}}$ \\ \hline 
\end{tabular} 
\caption{All possible boundary conditions for an M-theory end of the world brane with gauge algebra $G_2$. Each spectrum is accompanied by the factorized twelve-dimensional anomaly polynomial compatible with M-theory. In the last column, we also provide the four-form $G_4$ of M-theory to establish a precise connection, similar to the $E_8$ case by using~\eqref{bc}. All of these solutions can be obtained by simply restricting the spectrum of the $E_8$ Ho\v{r}ava-Witten wall to different $G_2$ subgroups, and as such, none of these boundary conditions is genuinely new.}
\label{table1}
\end{table}

It is worth mentioning that we have also found some curious solutions that have an appealing anomaly polynomial factorization, which is almost as the restricted Green-Schwarz factorization obeyed by the original $E_8$ gauge theory of Ho\v{r}ava-Witten or those found for $G_2$ in table \ref{table1}. An example is  
\begin{equation}(n_{\mathsf{1}}=102, n_{\mathsf{7}}=13, n_{\mathsf{14}}=2, n_{\mathsf{27}}=1),\label{nvsol}\end{equation}
for which the anomaly polynomial factorizes as
\begin{equation}-\frac{1}{768}(2\,c_{2,\mathsf{7}} + p_1)\left(\frac{2}{3}(6\,c_{2,\mathsf{7}} + p_1)^2 + \frac{7}{3}\,p_1^2 - 4\,p_2 \right).\label{nvpol}\end{equation}
The mismatch is in the fact that the anomaly polynomial cannot be written in terms of a single $G_4$ (at least two independent four-forms would be needed, in contradiction with M-theory), as well as in the coefficient of the pure gravitational anomaly. Whether or not this anomaly polynomial has a home somewhere in the non-supersymmetric string Landscape remains uncertain, but it certainly cannot be in the M-theory corner. For the interested reader, a few more similar solutions can be found in Appendix \ref{appA}.

All in all, we find no new purely chiral boundary conditions of M-theory; they all descend from the Ho\v{r}ava-Witten $E_8$ theory via different embeddings $G_2\rightarrow E_8$. One could suspect that the restriction that all matter fields have the same chirality may be too strong. To address this, we have also performed a scan allowing for non-chiral matter (effectively, this means that some of the $n_i$ coefficients may become negative). For computational reasons, we have restricted ourselves to representations of dimension less than $248$, as we did in the chiral case, even though there is no physical reason to do that now that both positive and negative chiralities are allowed. We have also let the coefficient $a$ in \eqref{eq:factAPG}, which controls the relation between the M-theory $G_4$-form and the Chern class of the $G_2$ bundle, to vary in the range
\begin{equation} a\,\in\, [0,100],\quad \text{and}\quad \vert n_i\vert \leq 200.\end{equation}
In this range, we find several solutions to the linear system of equations. It is already remarkable that there is any solution at all, since \eqref{eq:fermcounting} comprises a Diophantine system of equations over the integers.\footnote{Diophantine equations have also found application in the search for extensions of the Standard Model gauge group consistent with perturbative anomaly cancellation, see \cite{Appelquist:2002mw, Cui:2017juz,Costa:2019zzy, Allanach:2019uuu, Costa:2020dph, Allanach:2019gwp}, and for a more mathematical application, see \cite{Dobrescu:2020evn}} Among the solutions that we found, there is one which is somewhat special: Consider the spectrum \begin{equation}(n_{\mathsf{1}}=-27, n_{\mathsf{14}}=50, n_{\mathsf{27}}=-30, n_{\mathsf{77}}=5),\label{nvspecial}\end{equation} which, in our notation, corresponds to having 27 gauge neutral right-handed fermions, 30 right-handed fermions transforming in the $\mathsf{27}$ representation of $G_2$, along with 50 and 5 left-handed fermions transforming in the $\mathsf{14}$ and $\mathsf{77}$ representations of $G_2$ respectively. Using the data of table \ref{tableG2}, one can show that the anomaly polynomial is given by \eqref{eq:factAPG} with $a = 10$. This solution is minimal, in the sense that it only involves four different particle species to achieve anomaly cancellation (while there are five non-trivial anomaly constraints). It is the only solution with this property that we found in our scan. 

In short, the question ``does M-theory anomaly inflow allow for more general boundary conditions than the $E_8$-brane, once SUSY is dropped'' has an affirmative answer, and the most minimalistic such solution for $G_2$ group is given by \eqref{nvspecial}. The most natural question is whether this solution is physical, and corresponds to a novel $G_2$ end-of-the-world membrane in M theory, or if, on the contrary it is a spurious solution that does not have physical meaning. We will not solve this question here, but merely point out that since the coefficient $a=10$, and the smallest value that $c_2$ can have in the fundamental representation of $G_2$ is 2, we learn that, should such a brane exist, the small instanton limit of a worldvolume $G_2$ instanton has the same $C_7$ charge as 5 M5 branes (this is to be contrasted to the same calculation for $E_8$, for which the answer is 1 \cite{Ganor:1996mu}).

Again, it might be the case that this solution is spurious, and there is no end-of-the-world boundary condition for M-theory with $G_2$ gauge group. But it is fun to speculate what would the consequences be if it indeed existed. The obvious questions would be what would be its tension; since by assumption the brane is well-described by a $G_2$ gauge theory at low energies, the tension should not be large in Planck units. As mentioned above, there is also a small instanton transition, where 5 M5 branes together can ``puff up'' to become a $G_2$ gauge instanton. We expect that such a transition must be physically possible, since otherwise the $G_2$ instanton number would be a conserved charge, and therefore, a global symmetry (a similar argument predicts the transitions recently constructed in \cite{Anderson:2022bpo}). If so, at the small instanton point one would expect a non-supersymmetric $G_2$ version of the $E_8$ string \cite{Ganor:1996mu}, which would be very interesting to understand. Also, M2 branes should be able to end on the brane; the string at the intersection should have chiral degrees of freedom charged under $G_2$. We can figure out what these should be by compactifying on an inter to get a string whose global symmetry group would be $G_2\times G_2$. At low energies, where the string is described by a CFT, we therefore expect two copies of a $G_2$ Kac-Moody algebra at some level $\kappa$. Using anomaly inflow \cite{Kim:2019vuc}, we get
\begin{equation}\kappa=5,\end{equation}
and the corresponding current algebra central charge is
\begin{equation}c_{G_2\times G_2}=2\cdot \frac{\kappa\, \text{dim}(G_2)}{\kappa+2}=20.\label{cacc}\end{equation}
The left-moving central charge of the internal CFT is therefore at least 20. Together with the left and right-moving center of mass degrees of freedom of the string, the central charge is at least (30,10). Since perturbative ten-dimensional heterotic strings have an central charge of $(26,15)$, the $G_2$ string, if it exists, would be non-perturbative. This agrees with our arguments in the introduction and explains why it was not seen in the classification in \cite{BoyleSmith:2023xkd}.

We notice in passing that it is perhaps surprising that \eq{cacc} is integer. For a general level $\kappa$, the current algebra central charge is not an integer; in fact, the central charge is integer only for 
\begin{equation}\kappa=2,5\,\text{or}\, 12,\end{equation}
and for no other value of $\kappa$.  Integrability of the central charge is important for things like modular invariance, or for the CFT to be well defined as a Spin or oriented CFT. If the $G_2$ string was not corresponding to an actual string, there would be no reason to expect that the central charge is integer-valued. Of course, a fractional central charge might also be fixed by combining with another internal CFT, such as a minimal model; this is what happens, for instance, for the CHL string \cite{Kim:2019ths}. Nevertheless, the fact that the central charge is integral, and therefore one does not need to work hard to find an integer central charge, might be taken as circumstancial evidence for the existence of the string.

Another question is whether the interval compactification with $G_2\times G_2$ gauge group could have global anomalies. In \cite{Lee:2022spd}, it was show that the relevant spin bordism group vanishes; however, the relevant structure here is a twisted form of string bordism \cite{Basile:2023knk, Tachikawa:2024ucm}, so a global anomaly might remain. This is an interesting check that could be carried out in the future. At any rate, even if there is a global anomaly, it might be possible to cancel it with a topological Green-Schwarz mechanism, see \cite{Saito:2024iiu}.

In the following Subsections, we will perform similar analysis to this one for other exceptional groups. Let us advance that, with the restriction that the dimensions of all representations are below $248$, and allowing for any chirality, we have not found analogous (non)-chiral solutions to \eqref{eq:lsforsimplealg} and \eqref{eq:fermcounting}for any of the other exceptional groups; we did not extend the search to the infinite series of algebras $A,B,C,D$ due to computational complexity. Thus, among the exceptional groups, the solution \eqref{nvspecial} is quite unique. Whether this is because our search was somewhat limited or because \eqref{nvspecial} does correspond to a new solution in M-theory remains to be seen.

\subsection{\texorpdfstring{$F_4$}{F4} gauge group}
We continue our discussion with $F_4$. In this case, there are only three representations with dimension below $248$. We have found only one solution, listed in table~\ref{table2}, with all fields having the same chirality. This solution can be connected to the $E_8$ wall out one of the branchings we have already mentioned for $G_2$, namely $E_8 \supset G_2 \times F_4$. No new solutions appear when the same set of fields is considered but both chiralities are allowed (possibly, due to the smaller number of representations with dimensions below 248). Therefore, there are no new boundary conditions in this case either.

\begin{table}[!ht]
\centering
\renewcommand{\arraystretch}{2.5}
\begin{tabular}{c|c|c}
\hline
\multicolumn{3}{c}{$F_4$ gauge group} \\ \hline
& $(n_{\mathsf{1}}=14, n_{\mathsf{26}}=7, n_{\mathsf{52}}=1)$ & \\
$\mathcal{P}_{F_4}(R,F)=$ & $\displaystyle{-\frac{1}{96}\left(\frac{2}{3}\,c_{2,\mathsf{26}} + p_1\right) \left[\frac{1}{4}\left(\frac{2}{3}\,c_{2,\mathsf{26}} + p_1\right)^2 + \frac{1}{8}\,p_1^2 - \frac{1}{2}\,p_2 \right]}$ & $\displaystyle{G = \frac{1}{4} p_1 + \frac{1}{6} c_{2,{\mathsf{26}}}}$ \\ \hline 
\end{tabular} 
\caption{The spectrum for the only solution for  $F_4$, together with its anomaly polynomial and the  M-theory four-form. This solution also arises from a $F_4$ subgroup restriction of the $E_8$ brane}
\label{table2}
\end{table}

\subsection{\texorpdfstring{$E_6$}{E6} gauge group}
Next, we discuss $E_6$ gauge theory. The results are displayed in table \ref{table3}. Notice that all solutions feature fermions in a real representation; this is the case in dimensions 2 modulo 8 since a given representation $R$ is physically equivalent to its Majoranana conjugate \cite{green1988superstring}. 

\begin{table}[!ht]
\centering
\renewcommand{\arraystretch}{2.5}
\begin{tabular}{c|c|c}
\hline
\multicolumn{3}{c}{$E_6$ gauge group} \\ \hline
& $(n_{\mathsf{1}}=8, n_{\mathsf{27}}=3, n_{\mathsf{\overline{27}}}=3, n_{\mathsf{78}}=1)$ & \\
$\mathcal{P}_{E_6}(R,F)=$ & $\displaystyle{-\frac{1}{96}\left(\frac{2}{3}\,c_{2,\mathsf{27}} + p_1\right) \left[\frac{1}{4}\left(\frac{2}{3}\,c_{2,\mathsf{27}} + p_1\right)^2 + \frac{1}{8}\,p_1^2 - \frac{1}{2}\,p_2 \right]}$ & $\displaystyle{G = \frac{1}{4} p_1 + \frac{1}{6} c_{2,{\mathsf{26}}}}$ \\ \hline 
\end{tabular} 
\caption{Combinations of representations of $E_6$ with its respective anomaly polynomials.}
\label{table3}
\end{table}

We found a single solution (and we scanned for both positive and negative chirality representations), which can be found in table~\ref{table3} and is related to the branching $E_8 \supset E_6 \times SU(3)$. This is the unique embedding of $SU(3)$ in $E_8$, so again, we only find the (branching of the) Ho\v{r}ava-Witten solution.
Finally, note also that the anomaly polynomial of $F_4$ is the same as in this case. This is explained by the fact that $\mathsf{27} \rightarrow \mathsf{26} \oplus \mathsf{1}$ under $E_6 \supset F_4$.

\subsection{\texorpdfstring{$E_7$}{E7} gauge group}
Finally, we discuss $E_7$, for which we have only two nontrivial representations with dimension below $248$. Just like above, the only solution we found is the one coming from the branching  $E_7 \times SU(2) \subset E_8$, in table \ref{table4}.
\begin{table}[!ht]
\centering
\renewcommand{\arraystretch}{2.5}
\resizebox{\textwidth}{!}{
\begin{tabular}{c|c|c}
\hline
\multicolumn{3}{c}{$E_7$ gauge group} \\ \hline
& $(n_{\mathsf{1}}=3, n_{\mathsf{56}}=2, n_{\mathsf{133}}=1)$ & \\
$\mathcal{P}_{E_7}(R,F)=$ & $\displaystyle{-\frac{1}{96}\left(\frac{1}{3}\,c_{2,\mathsf{27}} + p_1\right) \left[\frac{1}{4}\left(\frac{1}{3}\,c_{2,\mathsf{27}} + p_1\right)^2 + \frac{1}{8}\,p_1^2 - \frac{1}{2}\,p_2 \right]}$ & $\displaystyle{G = \frac{1}{4} p_1 + \frac{1}{12} c_{2,{\mathsf{26}}}}$ \\ \hline 
\end{tabular} 
}
\caption{Combinations of representations of $E_6$ with its respective anomaly polynomials.}
\label{table4}
\end{table}

\subsection{Classical Lie algebras}
In the previous Subsections we focused on exceptional algebras, where trace identities lead to various simplifications. We now shift our attention to the more involved case of classical Lie algebras. Although we will not be able to solve the problem explicitly in full generality (since especially small algebras like $SU(2)$ or $Sp(2)$ have many representations of dimension below $248$, and many invariants to be computed), we will be able to place significant constraints in this case as well. 
We will start our search with the algebra $A_{n-1}$ i.e. $SU(n)$ group, then we proceed to analyze algebras $B_n, C_n, D_n$ for $n>2$.)

\subsubsection{\texorpdfstring{$SU(n)$}{SUn} algebra}\label{sunalg}
Here, we explicitly study the algebra $A_{n-1}$ for $n \geq 6$ and obtain solutions for the constraints \eqref{eq:lsforsimplealg} and  \eqref{eq:fermcounting}. For algebras with $n \leq 5$ the number of representations significantly increases as we go on lower $n$-values. Unfortunately, we have not been able to develop or find an efficient and systematic algorithm to compute indices for these lower order algebras\footnote{Except for $SU(2)$ algebra, which does not have indices greater than order two, this analysis could in principle be carried out. However, for computational reasons, we provide only a brief discussion in Subsection \ref{su2}.}, especially for sixth-order traces. By contrast, we were able to do so for $n\geq 6$, and the results may be found in Appendix~\ref{App:B}  (see \cite{Macfarlane:1999blf} for an algorithm to compute trace identities of the defining and adjoint representation matrices for these untreated lower-order algebras). Therefore, this work does not explicitly rule out algebras for $n \leq 5$. These lower-order algebras could be interesting to analyze, since the number of representations increases while the number of constraint equations decreases with $n$ due to accidental trace relations. We hope to return to this in the future.

We will now give a few more details and the additional traces present in this case and how have we dealt with them. Unlike the exceptional algebras just worked out, the classical algebras do not share the same simplification of trace identities. The sixth-order trace of an arbitrary representation $\mathsf{r}$ of $SU(n)$ is 
\begin{equation}\label{eq:sixthtrace}
\mathrm{tr}_{\mathsf{r}}\,F^6  = {\mathrm{u}}_{\mathsf{r}}^{(1)}\mathrm{tr}\,F^6 + {\mathrm{u}}_{\mathsf{r}}^{(3)}\mathrm{tr}\,F^4 \,\mathrm{tr}\,F^2  + {\mathrm{u}}_{\mathsf{r}}^{(5)}(\mathrm{tr}\,F^3)^2 +{\mathrm{u}}_{\mathsf{r}}^{(8)}(\mathrm{tr}\,F^2)^3,    
\end{equation}
with the coefficients ${\mathrm{u}}_{\mathsf{r}}^{(i)}$ matching those introduced before in the factorization of Chern characters (up to an irrelevant normalization). Similarly, for the fourth-order trace, we have
\begin{equation}
    \mathrm{tr}_{\mathsf{r}}\,F^4 = {\mathrm{u}}_{\mathsf{r}}^{(12)}\mathrm{tr}\,F^4 + {\mathrm{u}}_{\mathsf{r}}^{(14)}(\mathrm{tr}\,F^2)^2. 
\end{equation}
Our problem here amounts to exactly finding solutions to the homogeneous system \eqref{eq:lsforsimplealg}. Afterwards, we have to go through the inhomogeneous one, \eqref{eq:fermcounting}.

We have explicitly solved this problem for $A_{n \geq 5}$, ruling out all solutions except those that can be straightforwardly embedded in M-theory. These solutions are listed below, although their anomaly polynomials are not explicitly written down. To do this, one can follow the same reasoning as in the previous Section of exceptional algebras which was fully worked out, and the necessary Casimir invariants are found in Appendix \ref{App:B}. Note that, in table \ref{tab:Susolutions}, complex representations $\mathsf{r}$ always appear together with their complex conjugate $\bar{\mathsf{r}}$, as needed for them to give rise to a real one as required in ten dimensions. Unlike for exceptional groups, we have restricted ourselves to positive chirality representations only.
\begin{table}[!ht]
    \centering
    \renewcommand{\arraystretch}{1.8}
    \begin{tabular}{c|c}
        Solutions & M-theory embedding \\ \hline \hline
        $n_{\mathsf{80}} = 1$, $n_{\mathsf{84}, \overline{\mathsf{84}}} = 1$  & $E_8 \supset SU(9)$ \\ \hline
        $n_{\mathsf{1}} = 1$, $n_{\mathsf{8}, \overline{\mathsf{8}}} = 1$, $n_{\mathsf{28}, \overline{\mathsf{28}}} = 1$, $n_{\mathsf{56}, \overline{\mathsf{56}}} = 1$, $n_{\mathsf{63}} = 1$ & $E_8 \supset SU(9) \supset SU(8)$ \\ 
        $n_{\mathsf{1}} = 3$, $n_{\mathsf{28}, \overline{\mathsf{28}}} = 2$, $n_{\mathsf{63}} = 1$, $n_{\mathsf{70}} = 1$ & $E_8 \supset SO(16) \supset SU(8)$ \\ \hline
        $n_{\mathsf{1}} = 4$, $n_{\mathsf{7}, \overline{\mathsf{7}}} = 3$, $n_{\mathsf{21}, \overline{\mathsf{21}}} =2$, $n_{\mathsf{35}, \overline{\mathsf{35}}} = 1$, $n_{\mathsf{48}} = 1$ & $E_8 \supset SU(9) \supset SU(8) \supset SU(7)$ \\ \hline
        $n_{\mathsf{1}} = 11$, $n_{\mathsf{6}, \overline{\mathsf{6}}} = 6$, $n_{\mathsf{15}, \overline{\mathsf{15}}} =3$, $n_{\mathsf{20}} = 2$, $n_{\mathsf{35}} = 1$ & $E_8 \supset SU(9) \supset SU(8) \supset SU(7) \supset SU(6)$ \\ \hline
    \end{tabular}
    \caption{Solving the systems of equations~\eqref{eq:lsforsimplealg} and \eqref{eq:fermcounting} we get a set of solutions, leading to a factorized anomaly polynomial that can be embedded in the already known $E_8$ end of the world brane solution of M-theory~\cite{Horava:1995qa, Horava:1996ma}. Here, we have restricted to positive chirality solutions, and the notation $n_{\mathsf{r},\bar{\mathsf{r}}} = k, k \in \mathbb{Z}$ means that the combination $(\mathsf{r} \oplus \bar{\mathsf{r}})$ appears $k$-times in \eqref{eq:combiofreps} in agreement with reality of representations in 10d.}  
    \label{tab:Susolutions}
\end{table}

As is clear from the table, we found no solutions of positive chirality except the ones that embed in $E_8$. This mirrors what was happening in the exceptional groups.

\subsubsection{\texorpdfstring{$C_n$}{Cn} algebra}

We now explore the simple algebra $C_n$ for $n \geq 3$, including a summary of results and their embeddings within M-theory. We leave for future analysis the algebra $C_{n=2}$ for the same reason previously mentioned in Section \ref{sunalg} (there is a lot of representations). 

The set of homogeneous conditions that we have to solve comes from \eqref{eq:lsforsimplealg} with the linear equation for ${\mathrm{u}}_{\mathsf{r}_i}^{(5)}$ trivially satisfied. This can be seen from \eqref{eq:sixthtrace} since the $(\mathrm{tr}\,F^3)^2$ term is absent in $C_n$ algebras (this is also true for $B_n$ and $D_n$ algebras). Solutions that solve both \eqref{eq:lsforsimplealg} and \eqref{eq:fermcounting} are summarized in table \ref{tab:Cnsolutions} for algebras $C_3$ and $C_4$, again restricted to solutions of positive chirality only.  Only these cases are presented because we have not found any nontrivial solution for $C_n$ algebras with $n>4$. This matches the fact that there is no embedding of $C_{n>4}$ in $E_8$. The solutions we found for $C_3,C_4$ indeed correspond to $E_8$ branchings. We can actually give a direct argument that no solutions for $n>4$ can exist, just from looking at the $\text{tr}(F^4)$ term in the anomaly polynomial: All representations of dimension less than 248 have a positive contribution to this term, so with only chiral fields, it is impossible to cancel the obstruction.

\begin{table}[ht!]
    \centering
    \renewcommand{\arraystretch}{1.8}
    \begin{tabular}{c|c}
       Solutions & M-theory embedding \\ \hline \hline 
       $n_{\mathsf{1}} = 17, n_{\mathsf{6}} = 14, n_{\mathsf{14}} = 7, n_{\mathsf{14}'} = 2, n_{\mathsf{21}} = 1$  &  $E_8 \supset E_7 \supset \mathrm{Sp}(6)$ \\
       $n_{\mathsf{1}} = 1, n_{\mathsf{14}} = 2, n_{\mathsf{21}} = 1, n_{\mathsf{64}} = 2, n_{\mathsf{70}} = 1$  & $E_8 \supset SO(16) \supset SO(14) \supset \mathrm{Sp}(6)$ \\ \hline 
       $n_{\mathsf{1}} = 3, n_{\mathsf{8}} = 4, n_{\mathsf{27}} = 3, n_{\mathsf{36}} = 1, n_{\mathsf{48}} = 2$ & $E_8 \supset SU(9) \supset SU(8) \supset \mathrm{Sp}(8)$ \\ \hline
    \end{tabular}
    \caption{Solutions of $C_3$ and $C_4$ that can be embedded within M-theory. Correspondingly, the anomaly polynomial can be factorized as in the $E_8$ solution.}
    \label{tab:Cnsolutions}
\end{table}

A final comment (related to a remark in Subsection \ref{sec:G2}) is that for $C_3$ we  also found a solution where all fields have positive chirality and the anomaly polynomial factorizes, but not as it should to connect to M-theory. This is discussed in Appendix \ref{appA}. Again, we do not know if this has any physical meaning.

\subsubsection{Algebras \texorpdfstring{$B_n$}{Bn} and \texorpdfstring{$D_n$}{Dn}}

We have not studied this case thoroughly; however, at least for algebras $B_{n \geq 6}$ and $D_{n \geq 7}$ we can show that there are no solutions involving only fields with positive chirality beyond those that can be embedded in M-theory. To reach this conclusion, it is sufficient to examine the trace identities \cite{Okubo:1985qk}
\begin{align}\label{eq:sosymadj}
\begin{split}
\mathrm{tr}_{\mathsf{sym},\mathsf{a}}\,F^6 & = (2n \pm 32)\,\mathrm{tr}\,F^6 + 15\,(\mathrm{tr}\,F^4)(\mathrm{tr}\,F^2) \,, \\
\mathrm{tr}_{\mathsf{sym},\mathsf{a}}\,F^4 & = (2n \pm 8)\,\mathrm{tr}\,F^4 + 3\,(\mathrm{tr}\,F^2)^2 \,, \\
\mathrm{tr}_{\mathsf{sym},\mathsf{a}}\,F^2 & = (2n \pm 2)\,\mathrm{tr}\,F^2\,, \\
\end{split}
\end{align}
where $\mathsf{sym}$ and $\mathsf{a}$ stand for the symmetric and adjoint representations of rotation groups and for the cases where the $\mathsf{spin}$ representation is relevant i.e. less than $248$, we need \cite{green1988superstring} 
\begin{align}\label{eq:spin}
\begin{split}
\mathrm{tr}_{\mathsf{64}}\,F^6 & = 8\,\mathrm{tr}\,F^6 - \frac{15}{2}\,(\mathrm{tr}\,F^4)(\mathrm{tr}\,F^2) + \frac{15}{8}\,(\mathrm{tr}\,F^2)^3 \,, \\
\mathrm{tr}_{\mathsf{64}}\,F^4 & = -4\,\mathrm{tr}\,F^4 + 3\,(\mathrm{tr}\,F^2)^2 \,, \\
\mathrm{tr}_{\mathsf{64}}\,F^2 & = 8\,\mathrm{tr}\,F^2\,. \\
\end{split}
\end{align}
Up to certain normalization, equations \eqref{eq:sosymadj} and \eqref{eq:spin} give us the coefficients 
$\mathrm{u}_{\mathsf{r}_i}^{(1)}, \mathrm{u}_{\mathsf{r}_i}^{(3)}, \mathrm{u}_{\mathsf{r}_i}^{(8)},$
$\mathrm{u}_{\mathsf{r}_i}^{(12)}, \mathrm{u}_{\mathsf{r}_i}^{(14)}$ and $\mathrm{u}_{\mathsf{r}_i}^{(17)}$ needed to look for solutions for the systems \eqref{eq:lsforsimplealg} and \eqref{eq:fermcounting}. There are only two solutions to these equations, and they are precisely those that embed in the M-theory $E_8$-brane. In fact, both have appeared before:  one solution was worked out in Section \ref{sec:review} via the embedding $E_8 \supset SO(16)$. The other corresponds to $SO(14) \supset G_2$. Therefore, under the assumption of positive chirality this analysis allows us to rule out algebras $B_{n \geq 6}$ and $D_{n \geq 7}$. However, we have left out potentially interesting cases from the analysis, namely $SO(n)$ for $7\leq n \leq 12$, which we did not study explicitly due to the large number of representations involved and the lack of explicit formulae for some anomaly coefficients. All in all,  algebras deserve more attention, as their $\mathsf{spin}$ representations might yield new solutions that have no analog other algebras. We hope to return to this in the future. 

\subsubsection{\texorpdfstring{$SU(2)$}{SU2} algebra}\label{su2}
We have left the particular case of $SU(2)$ for last, since, even though it is the simplest Lie group, the aboundance of representations with dimensions blew 248 makes a complete analysis cumbersome. However in this particular case, the Casimir invariants are at most of order two, and they are known explicitly by the same techniques we applied to the exceptional groups. Thus to analyze this case we have to deal only with the following set of inhomogeneous equations
\begin{align}\label{su2ls}
\begin{split}
    \sum_{\mathsf{j}} n_{\mathsf{j}}\, T_2(\mathsf{j}) = 30\, a, \quad T_2(\mathsf{j})  & = \frac{2}{3} \mathsf{j}(\mathsf{j}+1)(2\mathsf{j}+1)\,,\\
    \sum_{\mathsf{j}} n_{\mathsf{j}}\, T_4(\mathsf{j})  = \frac{9}{4}\, a^2\, \quad T_4(\mathsf{j})  & = \frac{1}{15} \mathsf{j}(\mathsf{j}+1)(2\mathsf{j}+1)(3\mathsf{j}(\mathsf{j}+1)-1)\,, \\
    \sum_{\mathsf{j}} n_{\mathsf{j}} \, T_6(\mathsf{j})  = \frac{15}{32} \,a^3, \quad T_6(\mathsf{j})  & = \frac{1}{21} \mathsf{j}(\mathsf{j}+1)(2\mathsf{j}+1)[3\mathsf{j}(\mathsf{j}+1)(\mathsf{j}(\mathsf{j}+1) -1) + 1]\,,
\end{split}
\end{align}
where $T_2, T_4$, and $T_6$ are the second (normalized such that $T_2(\mathsf{1/2})=1$)-, fourth-, and sixth-order Dynkin indices \cite{mckay1981tables,Okubo:1981td, 10.1063:1.525670}, along with $\sum_{\mathsf{j}} n_{\mathsf{j}}\,\text{dim}(j) = 248$, where dim$(\mathsf{j}) = 2 \mathsf{j} +1$ for the dimension of the respective representation $\mathsf{j}$. One solution for this system is $(n_{\mathsf{0}}=8, n_{\mathsf{1}} = 28, n_{\mathsf{2}}=20, n_{\mathsf{3}} = 8)$ with $a = 32$ and corresponds to one way of embedding $ SU(2) \otimes SU(3)$ in $E_8$. Other direct solutions that correspond to the embeddings of M-theory are $E_8 \supset SU(2) \otimes E_7$ and $E_8 \supset SU(2)$. More generally, we have to solve a system of unknowns $n_{\mathsf{j}}, \mathsf{0} \leq \mathsf{j} \leq \mathsf{123}$ plus the variable $a$. We have not attempted to do this due to the computational complexity of the problem, but it should be remarked that we expect to find plenty, corresponding to the many different ways to embed $SU(2)$ into $E_8$. Furthermore, $SU(2)$ embeds on any other non-abelian group $G$, so by solving the non-chiral problem in full generality we will find avatars of all possible non-supersymmetric boundary conditions of M-theory that may exist (such as the $G_2$ solution in Subsection \ref{sec:G2}), for any Lie group $G$). This case therefore also constitutes a promising area for future research.

\section{Conclusions}\label{sec:conclusion}
M-theory admits an end of the world M9 brane with $E_8$ gauge fields and $\mathcal{N}=1$ supersymmetry, but we do not know whether this is the only possibility. In this paper we have taken the first steps towards identifying possible alternative end-of-the-world boundary conditions for M-theory, that satisfy the stringent anomaly cancellation criteria. Any such boundary condition other than the M9 brane would necessarily be nonsupersymmetric. 

We have written down the precise system of equations that encode the anomaly, and thoroughly explored the space of solutions where the end of the world brane contains a single simple factor $G$ and all chiral fermions have the same chirality. 

We have carried out a complete exploration of the algebras $G_2, F_4, E_6, E_7, C_{n \geq 3}$, finding solutions with an anomaly polynomial allowing for an embedding in M-theory. We have fully solved this problem in the case where the spectrum is assumed to be chiral, and found several solutions; however, the matter content of these can be traced back to branching rules of $E_8$ under those groups. Thus, they obey the factorization found in \cite{Witten:1996md}, but do not lead to a genuinely new solution -- they merely describe what happens when one takes the Horava-Witten brane and restricts the $E_8$ gauge group to a subgroup.
Algebras $A_{n \geq 5}, B_{n \geq 6}, D_{n \geq 7}$ can also be ruled out as candidates for new non-supersymmetric boundary conditions of M-theory, since we have shown that only possibilities or solutions that embed into the $E_8$ $\mathcal{N}=1$ theory exist. 

We also analyzed the problem with the assumption of a fully chiral spectrum in other cases where we could not be exhaustive. For the algebra $C_2$, there exist certainly solutions with embedding within M-theory since this is a subgroup of $E_8$. For the algebras $A_{n \leq 5}, B_{n \geq 6}, D_{n \leq 7}$, there is room to find solutions that can be embedded in M-theory as well; however, we did not perform an exhaustive search. Although one could think that these algebras are much easier to deal with due to the lack of genuine sixth-order Casimir invariants \cite{10.1063:1.525670},  the larger the number of representations that need to be considered increases the complexity of the problem; so some chiral solutions not coming from the $E_8$ could exist for these groups, in principle. 

We also find some solutions with a purely chiral spectrum, particularly for $G_2$ and $C_3$, where the anomaly polynomial factorizes, but not in a way compatible with M-theory. Whether this is a mere curiosity, or these solutions will find some place in the non-supersymmetric Landscape of string theory is a question that remains open.

Perhaps our most interesting result comes from the case of the exceptional groups $G_2,F_4,E_6,E_7$, where we extended our search to fermion spectra of both chiralities. We did not find any new solutions for $E_6,E_7$ and $F_4$, but for $G_2$ we found a non-chiral spectrum, involving just four different representations (the singlet $\mathsf{1}$, the adjoint $\mathsf{14}$, the $\mathsf{27}$, and the $\mathsf{77}$), whose anomaly polynomial factorizes exactly in the manner required to match M-theory anomaly inflow. In this case, there are five non-trivial anomaly cancellation conditions taking the form of Diophantine equations, solved simultaneously by only four matter fields. The resulting matter spectrum is extremely interesting, and one is left to wonder whether it could correspond to a non-supersymmetric boundary of M-theory. If so, its small instanton transition (where a worldvolume gauge instanton becomes five M5 branes) would describe a non-supersymmetric CFT in six dimensions. The worldvolume content of the associated ``$G_2\times G_2$'' heterotic-like string, obtained from M2's suspended between two $G_2$ boundaries, would have a current algebra at level 5 and an internal CFT central charge above 20, so if it indeed exists it would be a non-perturbative string.  Perhaps the most interesting concrete research question coming out of the simple analysis in this note is to find out whether this $G_2$ end of the world brane belongs to the Landscape or the Swampland. If it did exist, it would open up a whole new swath of the non-supersymmetric string Landscape, including the possibility of mixed $E_8\times G_2$ compactifications where the $G_2$ brane constitutes a SUSY-breaking dark sector which is only weaky transmitted to the $E_8$ brane\footnote{We thank A. Uranga for raising this interesting possibility.}. More generally, it would be good to understanding the full space of non-chiral space of solutions, and elucidate whether exotic boundary conditions of M-theory actually exist.
\vspace{0.3cm}

\textbf{Acknowledgements:} We are indebted to Bobby Samir Acharya, Fernando Quevedo, and Angel Uranga for useful discussions and comments on the manuscript.
The authors thank CERN for hospitality during the Strings 2024 conference, in which this work was very much advanced. MM thanks as well Harvard University and its Swampland Initiative for hosting and providing a stimulating environment where parts of this work were completed. LZ would also like to thank H. Garc\'ia-Compe\'an for helpful conversations and all the support and encouragement during completion of this work. 
MM is currently supported by the RyC grant RYC2022-037545-I from the AEI and was supported
by an Atraccion del Talento Fellowship 2022- T1/TIC-23956 from Comunidad de Madrid in
the early stages of this project.  The authors thank the
Spanish Research Agency (Agencia Estatal de Investigacion) through the grants IFT Centro de Excelencia Severo Ochoa CEX2020-001007-S and PID2021-123017NB-I00, funded by
MCIN/AEI/10.13039/501100011033 and by ERDF A way of making Europe. LZ extends his deepest gratitude to Swamplandia workshop and its early-career scholarship, as well as to the IFT for its generous hospitality, where this work was initiated. His work was supported by grant 941366 from SECIHTI (known as CONACyT before 2025). He also thanks the Programa de Becas Elisa Acu\~na through the grants FIS-EX-2023-699 and FIS-ES-2023-701, which contributed to the completion of this work. LZ is currently supported by a postdoctoral fellowship from NYUAD.

\appendix

\section{Solutions with an unrestricted Green-Schwarz factorization}\label{appA}
The search for end-of-the-world branes of M-theory requires a matter content that is heavily constrained by the bulk anomaly inflow. As we saw in Section \ref{sec:searchbc}, the anomaly polynomial of the matter content must admit a specific Green-Schwarz factorization; otherwise the bulk-boundary system cannot be consistently defined. However, while exploring boundary conditions for M-theory with gauge group $G_2$, we found certain spectra whose anomaly polynomials satisfy a Green-Schwarz factorization, which however is inconsistent with M-theory anomnaly inflow.  In this Appendix, we provide a table with a summary of these anomaly polynomials that cannot be embedded within M-theory. 

\begin{table}[!ht]
\centering
\renewcommand{\arraystretch}{2}
\begin{tabular}{c|c|c}
\multirow{2}{*}{$G_2$} & $(n_{\mathsf{1}}=152, n_{\mathsf{14}}=3, n_{\mathsf{27}}=2)$ & \\
&  $\displaystyle{ -\frac{1}{768}(2\,c_{2,\mathsf{7}} + p_1)\left(\frac{2}{5}(10\,c_{2,\mathsf{7}} + p_1)^2 + \frac{13}{5}\,p_1^2 - 4\,p_2 \right)}$ & \\ \hline
\multirow{2}{*}{$G_2$} & $(n_{\mathsf{1}}=102, n_{\mathsf{7}}=13, n_{\mathsf{14}}=2, n_{\mathsf{27}}=1)$ & \\
& $\displaystyle{ -\frac{1}{768}(2\,c_{2,\mathsf{7}} + p_1)\left(\frac{2}{3}(6\,c_{2,\mathsf{7}} + p_1)^2 + \frac{7}{3}\,p_1^2 - 4\,p_2 \right)}$ & \\ \hline
\multirow{2}{*}{$\mathrm{Sp}(6)$} & $(n_{\mathsf{1}}=64, n_{\mathsf{6}}=14, n_{\mathsf{14}'}=3, n_{\mathsf{64}}=1)$ & \\
& $\displaystyle{ -\frac{1}{768}(4\,c_{2,\mathsf{7}} + p_1)\left(\frac{8}{10}(10\,c_{2,\mathsf{7}} + p_1)^2 + \frac{11}{5}\,p_1^2 - 4\,p_2 \right)}$ & \\
\end{tabular} 
\caption{Combinations of representations with their respective anomaly polynomials which are not directly related to branching rules but, with an appealing anomaly polynomial factorization.}
\label{tableg2sp6}
\end{table}

These solutions are presented in table \ref{tableg2sp6}. Two of them correspond to $G_2$ spectra, while the other is characterized by an $\text{Sp}(2n)$ matter content. However, it is clear that the factorizations we found are incompatible with the M-theory anomaly polynomial \eqref{eq:topcouMtheory4}. Whether or not these solutions have a meaning elsewhere in the string Landscape is a question that remains open. 

\section{Anomaly polynomials and indices}\label{App:B}
We start with a lightning review of perturbative anomaly cancellation; for more details, see \cite{Alvarez-Gaume:1986ghj}. In the study of perturbative anomalies of a chiral fermion theory in a $d$-dimensional Spin manifold $M$ the Atiyah-Singer index theorem plays an important role. This theorem allows us to determine the index of a Dirac operator, defined on a $(d+2)$ manifold $W$ with the appropriate structure extended from $d$-dimensions, in terms of geometrical data related to the tangent bundle and possibly a principal $G$-bundle if there is matter transforming under some representation of $G$.
Concretely, the index theorem of the Dirac operator of chiral spin-1/2 fermion\footnote{For a spin-3/2 fermion the index density changes as $$\hat{\mathrm{A}}(R)(\mathrm{ch}(R) - 1)$$ where we have subtracted a pure a gravitational term that accounts for ghosts \cite{Alvarez-Gaume:1984zlq} and $F \rightarrow R$.} says that~\cite{Atiyah:1968mp, Atiyah:1963zz}
\begin{equation}\label{eq:IndexTh}
    \mathrm{Index}(\mathcal{D}_{d+2}) = \int_{W} \left. \hat{\mathrm{A}}(R)\,\mathrm{ch}_{\mathsf{r}}(F)\right|_{d+2}\,,
\end{equation}
where the $\hat{\mathrm{A}}(R)$ is the $\mathrm{A}$-roof or Dirac genus given in terms of traces of powers of the Riemann curvature two-form $R$, i.e. $\mathrm{tr}\,R^n$, $\mathrm{ch}_{\mathsf{r}}(F) = \mathrm{tr}_{\mathsf{r}} \exp\left(\frac{\mathsf{i}}{2 \pi}\,F \right)$ is the Chern character in terms of the field strength associated to a connection $A$ with $\mathrm{tr}_{\mathsf{r}}$ a trace evaluated in some representation $\mathsf{r}$ of the Lie algebra of $G$. When we write $\mathrm{tr}$ without a subscript it means a trace evaluated in some reference representation of the group. The notation $\hat{\mathrm{A}}(R)\,\mathrm{ch}_{\mathsf{r}}(F)|_{d+2}$ for the index density means that only terms of degree $d+2$ survive.

Following standard conventions in mathematical literature, the $A$-roof Dirac genus can be expressed as
\begin{equation}\label{eq:Aroof}
    \hat{\mathrm{A}}(R) = 1 - \frac{1}{24} p_1 + \frac{1}{5760}(7\,p_1^2 - 4\,p_2) - \frac{1}{967680}(31\,p_1^3 - 44\,p_1\,p_2 + 16\,p_3) + \cdots
\end{equation}
where we have defined $p_i \equiv p_i(R)$ and correspond to the $4i$-th Pontryagin classes.

On the other hand, the Chern character is determined by 
\begin{align}
    \begin{split}
        \mathrm{ch}_{\mathsf{r}}(F) & = \sum_l \frac{1}{l!}\,\mathrm{tr}_{\mathsf{r}}\left(\frac{\mathsf{i}\,F}{2 \pi} \right)^l\,, \\
        & = \sum_l \mathrm{ch}_{l,\mathsf{r}}(F)
    \end{split}
    \end{align}
where $F \equiv \mathsf{i}\, F/2 \pi$ and the $l$-th Chern character will be denoted as $\mathrm{ch}_{l,\mathsf{r}}(F) = \mathrm{ch}_{l,\mathsf{r}}$ rather than specifically in terms of traces of powers of the field strength. These characters can be related to Chern classes $c_{i,\mathsf{r}}(F)$, see, e.g. \cite{nakahara2003geometry}.

We  will focus on simple algebras in this appendix, so the problem of anomaly factorization demands dealing with the following trace identities for an arbitrary representation $\mathsf{r}$ of a simple gauge group
\begin{align}
\begin{split}
\label{eq:traces}
\mathrm{tr}_{\mathsf{r}}\,F^6 & = {\mathrm{u}}_{\mathsf{r}}^{(1)}\mathrm{tr}\,F^6 + {\mathrm{u}}_{\mathsf{r}}^{(2)}(\mathrm{tr}\,F^4)(\mathrm{tr}\,F^2) + {\mathrm{u}}_{\mathsf{r}}^{(3)}(\mathrm{tr}\,F^2)^3 + {\mathrm{u}}_{\mathsf{r}}^{(4)}(\mathrm{tr}\,F^3)^2 \, \\ 
\mathrm{tr}_{\mathsf{r}}\,F^4 & = {\mathrm{u}}_{\mathsf{r}}^{(5)}\mathrm{tr}\,F^4 + {\mathrm{u}}_{\mathsf{r}}^{(6)}(\mathrm{tr}\,F^2)^2 \,,  \\ 
\mathrm{tr}_{\mathsf{r}}\,F^2 & = {\mathrm{u}}_{\mathsf{r}}^{(7)}\mathrm{tr}\,F^2\,, 
\end{split}
\end{align}
where the coefficients ${\mathrm{u}}_{\mathsf{r}}^{(i)}$ are related to eigenvalues of Casimir invariants or equivalently to $2n^{\mathrm{th}}$ (modified) order index. Our main task is to compute those coefficients. Note also that the coefficient ${\mathrm{u}}_{\mathsf{r}}^{(4)}$ is only present for $A_{n-1}$ algebras with $n>2$. For exceptional algebras, these coefficients can be found in \cite{Okubo:1981td, vanRitbergen:1998pn} which we summarize in table \ref{tableG2}.  

It is quite convenient to determine the coefficients ${\mathrm{u}}_{\mathsf{r}}^{(i)}$ for $A_{n-1}$ algebras using the Chern character and useful properties that obey under direct sum and tensor product of representations, namely 
\begin{align}
    \ch_{\mathsf{r}_1 \oplus \mathsf{r}_2}(F) & = \ch_{\mathsf{r}_1}(F) + \ch_{\mathsf{r}_2}(F)\,,\\
    \ch_{\mathsf{r}_1 \otimes \mathsf{r}_2}(F) & = \ch_{\mathsf{r}_1}(F) \, \ch_{\mathsf{r}_2}(F)\,.
\end{align}
Afterwards, we can use the information obtained for $A_{n-1}$ to analyze the algebras $B_n, C_n$, and $D_n$ using branching rules of representations. Unfortunately, we find that this method is only efficient for algebras with $n \geq 6$. 

What makes these properties of the Chern character particularly useful for representations of $SU(n)$ is the fact that we can express Chern characters of symmetric and anti-symmetric representations in terms of a defining representation. Eventually, we will manage to determine the group-theoretical coefficients ${\mathrm{u}}_{\mathsf{r}}^{(i)}$ of the traces above. This is done using the following identities~\cite{Schellekens:1986xh} 
\begin{align}\label{eq:antisymm}
    \sum_{m = 0}^{\infty}t^m \mathrm{ch}_{[m]}(F) & = \mathrm{det}\left(1 + t \, \exp \left(\mathsf{i}\frac{F}{2 \pi} \right)\right)\,, \\ \label{eq:symm}
    \sum_{m = 0}^{\infty}t^m \mathrm{ch}_{(m)}(F) & = \mathrm{det}\left(1 - t \,\exp \left(\mathsf{i}\frac{F}{2 \pi} \right)\right)^{-1}\,,
\end{align}
where $[m]$ and $(m)$ denote (anti)-symmetrized representations where the Chern character is evaluated. These are irreducible representations for $SU(n)$ using as reference representation its fundamental representation. For instance, if we denote by $V$ the fundamental of $SU(n)$, then the representation $[m]$ can be thought of as an element of the exterior algebra $\Lambda(V) = \oplus_m \Lambda^m(V)$. Therefore, the right-hand side of \eqref{eq:antisymm} and \eqref{eq:symm} is valued in the fundamental of $SU(n)$. This allows us to determine the corresponding Chern characters since we know how to deal with determinants 
\begin{align}
    \mathrm{det}\left(1 + t \, \exp \left(\mathsf{i}\frac{F}{2 \pi} \right)\right) & = \prod_{k=1}^{\infty} \exp\left[-\frac{(-t)^k}{k}\mathrm{ch}(k\,F) \right]\,, \\
    \mathrm{det}\left(1 - t \,\exp \left(\mathsf{i}\frac{F}{2 \pi} \right)\right)^{-1} & = \prod_{k=1}^{\infty} \exp\left[\frac{t^k}{k}\mathrm{ch}(k\,F) \right]\,,
\end{align}
where, as in the main text, $\ch(F) = \mathrm{tr}(\mathsf{i} \frac{F}{2 \pi})$ denotes the Chern character evaluated in the fundamental or defining representation. With this, one can show that
\begin{align}
    \ch_{[2]}(F) & = \frac{1}{2}\ch^2(F) - \frac{1}{2}\ch(2F)\,,\\
    \ch_{[3]}(F) & = \frac{1}{6} \ch^3(F) - \frac{1}{2}\ch(2F)\,\ch(F) + \frac{1}{3}\ch(3F)  \,,\\
    \ch_{[4]}(F) & = \frac{1}{24}\ch^4(F) - \frac{1}{4} \ch^2(F)\,\ch(2F) + \frac{1}{8}\ch^2(2F) + \frac{1}{3}\ch(F)\,\ch(3F) - \frac{1}{4}\ch(4F)\,.
\end{align}
From this, one can obtain basic trace identities with the corresponding index coefficients (up to some normalization) plus product of basic traces as well as the dimension of the representation $[\mathsf{m}]$. In fact, we have computed the indices for the representations $[\mathsf{2}], [\mathsf{3}], [\mathsf{4}], (\mathsf{2})$ and $(\mathsf{3})$ for the classical Lie algebra $A_{n-1}$, which are useful for our purposes. They are related to the coefficients ${\mathrm{u}}_{\mathsf{r}}^{(i)}$ and given by 
\begin{align}\label{eq:indices}
\begin{split}
    I_k([2]) & = n-2^{k-1} \, \\
    I_k([3]) & = \frac{1}{2}(n^2 - (1+2^k)n +2\cdot 3^{k-1})\, \\
    I_k((2)) & = n+2^{k-1} \, \\
    I_k((3)) & = \frac{1}{2}(n^2 + (1+2^k)n +2\cdot 3^{k-1})\, \\
    I_k([4]) & = \frac{1}{6}(n^3 - 3(2^{k-1}+1)n^2 + 2(3 \cdot 2^{k-2} +3^k +1)n - 6 \cdot 4^{k-1}) \,, \\
\end{split}
\end{align}
where $k \in \mathbb{Z}$. Nonetheless, we need the complete factorization of the corresponding traces for each of the representations. A cumbersome but straightforward computation gives us 
\begin{align}
\begin{split}
    \mathrm{tr}_{[2],(2)}F^6 & = (n \mp 2^5) \mathrm{tr}F^6 + 15\,\mathrm{tr}F^4\,\mathrm{tr}F^2 + 10\,(\mathrm{tr}F^3)^2\,, \\
    \mathrm{tr}_{[2],(2)}F^4 & = (n \mp 2^3) \mathrm{tr}F^4 + 3\,(\mathrm{tr}F^2)^2\,, \\
    \mathrm{tr}_{[2], (2)}F^2 & = (n \mp 2)\mathrm{tr}F^2\,,
\end{split}
\end{align}
where the upper sign is for the antisymmetric $[2]$ and the lower sign is for the symmetric $(2)$ representations. Whereas, for the representations $[3]$ and $(3)$, we have determined that 
\begin{align}
\begin{split}
    \mathrm{tr}_{[3],(3)}F^6 & = \frac{1}{2}(n^2 \mp 65n + 486)\mathrm{tr}F^6 + 15(n \mp 10)\mathrm{tr}F^4\,\mathrm{tr}F^2 
    + 10(n \mp 8)(\mathrm{tr}F^3)^2 + 15 (\mathrm{tr}F^2)^3\,, \\
    \mathrm{tr}_{[3],(3)}F^4 & = \frac{1}{2}(n^2 \mp 17n + 54) \mathrm{tr}F^4 + 3(n \mp 4)(\mathrm{tr}F^2)^2\,,\\[7pt]
    \mathrm{tr}_{[3],(3)}F^2 & = \frac{1}{2}(n^2 \mp 5n + 6)\mathrm{tr}F^2\,.
\end{split}
\end{align}
Finally, for the representation $[4]$, we find that
\begin{align}
\begin{split}
    \mathrm{tr}_{[4]}F^6 = & \frac{1}{6}(n^3 - 99n^2 + 1556n - 6144) \mathrm{tr}F^6 + \frac{15}{2}(n^2 - 21n + 92) \mathrm{tr}F^4\,\mathrm{tr}F^2 \\
     + & 5(n^2 - 17n + 68)(\mathrm{tr}F^3)^2 + 15(n - 6) (\mathrm{tr}F^2)^3 \,, \\
    \mathrm{tr}_{[4]}F^4 = & \frac{1}{6}(n^3 - 27n^2 + 188n - 384) \mathrm{tr}F^4 + \frac{3}{2}(n^2 - 9n +20)(\mathrm{tr}F^2)^2\,,\\
    \mathrm{tr}_{[4]}F^2 = & \frac{1}{6}(n^3 - 9n^2 + 26n - 24)\mathrm{tr}F^2\,.
\end{split}
\end{align}
Nevertheless, this is not enough since for algebras with $n \leq 9$ there are additional representations beyond those already worked out in this appendix, which play a role in the search for M-theory boundary conditions carried out in the main part of this work. Fortunately for us, we can use the fact that the tensor product of representations is the direct sum of, in general, reducible representations, namely
\begin{equation}
    \mathsf{r}_1 \otimes \mathsf{r}_2 = \oplus_i n_i \,\mathsf{r}_i 
\end{equation}
under which 
\begin{equation}\label{eq:produccharact} 
    \mathrm{ch}_{\mathsf{r}_1}(F) \, \mathrm{ch}_{\mathsf{r}_2}(F) = \sum_i n_i  \mathrm{ch}_{\mathsf{r}_i}(F).
\end{equation}
Thus, by computing the tensor product of two known characters for representations $\mathsf{r}_1$ and $\mathsf{r}_2$ we can calculate the character of an unknown representation by subtracting what we already know from the right-hand side of~\eqref{eq:produccharact}.

Consider the next simple example. Let $\mathsf{16} \otimes \mathsf{16}$ be the tensor product of the fundamental representation of $SU(16)$ which can be decomposed as follows
\begin{equation}
    \mathsf{16} \otimes \mathsf{16} = \mathsf{120} \oplus \mathsf{136},
\end{equation}
By using~\eqref{eq:produccharact} and~\eqref{eq:indices} for the representation $[2] = \mathsf{120}$ we find that 
\begin{equation}
    \mathrm{ch}_{\mathsf{136} = [3]}(F) = 136 + \frac{18}{2!} \mathrm{tr}(F)^2 + \frac{20}{3!} \mathrm{tr}(F)^3 + \frac{24}{4!} \mathrm{tr}(F)^4 + \frac{32}{5!} \mathrm{tr}(F)^5 + \frac{48}{6!} \mathrm{tr}(F)^6 + \cdots
\end{equation}
where $\cdots$ represents higher-order traces and products of lower ones.

It is also useful to recall the trace identities for the adjoint representation of $SU(n)$ \cite{Okubo:1985qk} 
\begin{align}\label{eq:suadjoint}
\begin{split}
\mathrm{tr}_{\mathsf{a}}\,F^6 & = 2n\,\mathrm{tr}\,F^6 + 30\,(\mathrm{tr}\,F^4)(\mathrm{tr}\,F^2)-20(\mathrm{tr}\,F^3)^2 \,, \\
\mathrm{tr}_{\mathsf{a}}\,F^4 & = 2n\,\mathrm{tr}\,F^4 + 6\,(\mathrm{tr}\,F^2)^2 \,, \\
\mathrm{tr}_{\mathsf{a}}\,F^2 & = 2n\,\mathrm{tr}\,F^2\,. \\
\end{split}
\end{align}
Ultimately, this abstract analysis boils down to the information gathered together in tables \ref{tablesu9}, \ref{tablesu8}, \ref{tablesu7}, and \ref{tablesu6}.

\begin{table}[!ht]
\centering
\renewcommand{\arraystretch}{2.15}
\begin{tabular}{p{1cm}||p{2cm}p{2cm}p{2cm}p{2cm}p{2cm}}
\hline
\multicolumn{6}{c}{Exceptional Algebras} \\ \hline \hline
$G$ & $\mathsf{r}$ & $\mathrm{u}_{\mathsf{r}}^{(1)}$ & $\mathrm{u}_{\mathsf{r}}^{(4)}$ & $\mathrm{u}_{\mathsf{r}}^{(6)}$ & $\mathrm{u}_{\mathsf{r}}^{(7)}$ \\ \hline 
\multirow{8}{*}{$G_2$}& $\mathsf{7}$ & $1$ & $-$ & $\displaystyle{\frac{1}{4}}$ & $1$ \\
                      & $\mathsf{14}$ & $-26$ & $\displaystyle{\frac{15}{4}}$ & $\displaystyle{\frac{5}{2}}$ & $4$ \\
                      & $\mathsf{27}$ & $39$ & $\displaystyle{\frac{15}{4}}$ & $\displaystyle{\frac{27}{4}}$ & $9$ \\
                      & $\mathsf{64}$ & $-208$ & $75$ & $38$ & $32$ \\
                      & $\mathsf{77}$ & $494$ & $\displaystyle{\frac{315}{4}}$ & $\displaystyle{\frac{121}{2}}$ & $44$ \\
                      & $\mathsf{77}'$ & $-1235$ & $\displaystyle{\frac{1275}{4}}$ & $\displaystyle{\frac{385}{4}}$ & $55$ \\
                      & $\mathsf{182}$ & $3666$ & $\displaystyle{\frac{2925}{4}}$ & $\displaystyle{\frac{663}{2}}$ & $156$ \\
                      & $\mathsf{189}$ & $-456$ & $735$ & $270$ & $144$ \\ \hline
\multirow{2}{*}{$F_4$} & $\mathsf{26}$ & $1$ & $-$ & $\displaystyle{\frac{1}{12}}$ & 1 \\
					  & $\mathsf{52}$ & $-7$ & $\displaystyle{\frac{5}{36}}$ & $\displaystyle{\frac{5}{12}}$ & $3$ \\ 
\hline  
\multirow{2}{*}{$E_6$} & $\mathsf{27}$ & $1$ & $-$ & $\displaystyle{\frac{1}{12}}$ & $1$ \\
					   & $\mathsf{78}$ & $-6$ & $\displaystyle{\frac{5}{36}}$ & $\displaystyle{\frac{1}{2}}$ & $4$ \\
\hline
\multirow{2}{*}{$E_7$} & $\mathsf{56}$ & $1$ & $-$ & $\displaystyle{\frac{1}{24}}$ & $1$ \\
					   & $\mathsf{133}$ & $-2$ & $\displaystyle{\frac{5}{288}}$ & $\displaystyle{\frac{1}{6}}$ & $3$ \\
\hline
\end{tabular}
\caption{This table summarizes representations and their trace-identity indices for dimension less than $\mathsf{248}$ for exceptional algebras. The coefficients in \eqref{eq:traces} that do not do not appear in this table are trivial. In the main text, this coefficients are identified with $\mathrm{u}_{\mathsf{r}}^{(1)}$,  $\mathrm{u}_{\mathsf{r}}^{(8)}$,  $\mathrm{u}_{\mathsf{r}}^{(14)}$, and  $\mathrm{u}_{\mathsf{r}}^{(17)}$ respectively.}
\label{tableG2}
\end{table}

\begin{table}[!ht]
\centering
\renewcommand{\arraystretch}{2.3}
\resizebox{\textwidth}{!}{\small
\begin{tabular}{p{1cm}||p{1.5cm}|p{1.2cm}|p{1.2cm}|p{1.2cm}|p{1.2cm}|p{1.2cm}|p{1.2cm}|p{1.2cm}|}
\hline
\multicolumn{9}{c}{Classical Lie Algebras} \\ \hline \hline
$G$ & $\mathsf{r}$ &  $\mathrm{u}_{\mathsf{r}}^{(1)}$ & $\mathrm{u}_{\mathsf{r}}^{(2)}$ &  $\mathrm{u}_{\mathsf{r}}^{(3)}$ &  $\mathrm{u}_{\mathsf{r}}^{(4)}$ &  $\mathrm{u}_{\mathsf{r}}^{(5)}$ &  $\mathrm{u}_{\mathsf{r}}^{(6)}$ & $\mathrm{u}_{\mathsf{r}}^{(7)}$ \\ \hline 
\multirow{8}{*}{$SU(9)$}& $\sf{9}, \overline{9}$ & $1$ & $-$ & $-$ & $-$ & $1$ & $-$ & $1$ \\
                      & $\sf{36}, \overline{36}$ & $-23$ & $15$ & $-$ & $10$ & $1$ & $3$ & $7$ \\
                      & $\sf{45}, \overline{45}$ & $41$ & $15$ & $-$ & $10$ & $17$ & $3$ & $11$ \\
                      & $\sf{80}$ & $18$ & $30$ & $-$ & $-20$ & $18$ & $6$ & $18$ \\
                      & $\sf{84}, \overline{84}$ & $-9$ & $-15$ & $15$ & $10$ & $-9$ & $15$ & $21$ \\
                      & $\sf{126}, \overline{126}$ & $95$ & $-120$ & $45$ & $-20$ & $-25$ & $30$ & $35$ \\
                      & $\sf{165}, \overline{165}$ & $576$ & $285$ & $15$ & $170$ & $144$ & $39$ & $66$ \\
                      & $\sf{240}, \overline{240}$ & $-162$ & $270$ & $30$ & $180$ & $54$ & $54$ & $78$ \\ \hline
\end{tabular}}
\caption{Representations and their trace-identity indices for $A_8$. In the main text, they are identified with $\mathrm{u}_{\mathsf{r}}^{(1)}$, $\mathrm{u}_{\mathsf{r}}^{(3)}$, $\mathrm{u}_{\mathsf{r}}^{(8)}$, $\mathrm{u}_{\mathsf{r}}^{(5)}$, $\mathrm{u}_{\mathsf{r}}^{(12)}$, $\mathrm{u}_{\mathsf{r}}^{(14)}$, and $\mathrm{u}_{\mathsf{r}}^{(17)}$.}
\label{tablesu9}
\end{table}

\begin{table}[!ht]
\centering
\renewcommand{\arraystretch}{2.3}
\resizebox{\textwidth}{!}{\small
\begin{tabular}{p{1cm}||p{1.5cm}|p{1.2cm}|p{1.2cm}|p{1.2cm}|p{1.2cm}|p{1.2cm}|p{1.2cm}|p{1.2cm}|}
\hline \hline
$G$ & $\mathsf{r}$ & $\mathrm{u}_{\mathsf{r}}^{(1)}$ & \centering $\mathrm{u}_{\mathsf{r}}^{(2)}$ & \centering $\mathrm{u}_{\mathsf{r}}^{(3)}$ & \centering $\mathrm{u}_{\mathsf{r}}^{(4)}$ & \centering $\mathrm{u}_{\mathsf{r}}^{(5)}$ & \centering $\mathrm{u}_{\mathsf{r}}^{(6)}$ & \hspace{0.25cm} $\mathrm{u}_{\mathsf{r}}^{(7)}$ \\ \hline 
\multirow{9}{*}{$SU(8)$}& $\sf{8}, \overline{8}$ & $1$ & $-$ & $-$ & $-$ & $1$ & $-$ & $1$ \\
                      & $\sf{28}, \overline{28}$ & $-24$ & $15$ & $-$ & $10$ & $-$ & $3$ & $6$ \\
                      & $\sf{36}, \overline{36}$ & $40$ & $15$ & $-$ & $10$ & $16$ & $3$ & $10$ \\
                      & $\sf{56}, \overline{56}$ & $15$ & $-30$ & $15$ & $-$ & $-9$ & $12$ & $15$ \\
                      & $\sf{63}$ & $16$ & $30$ & $-$ & $-20$ & $16$ & $6$ & $16$ \\
                      & $\sf{70}$ & $80$ & $-90$ & $30$ & $-20$ & $-16$ & $18$ & $20$ \\
                      & $\sf{120}, \overline{120}$ & $535$ & $270$ & $15$ & $160$ & $127$ & $36$ & $55$ \\
                      & $\sf{168}, \overline{168}$ & $-179$ & $240$ & $30$ & $160$ & $37$ & $48$ & $61$ \\
                      & $\sf{216}, \overline{216}$ & $-165$ & $210$ & $45$ & $-$ & $27$ & $60$ & $75$ \\                      
                      \hline
\end{tabular}
}
\caption{Representations and their trace-identity indices for $A_7$. In the main text, they are identified with $\mathrm{u}_{\mathsf{r}}^{(1)}$, $\mathrm{u}_{\mathsf{r}}^{(3)}$, $\mathrm{u}_{\mathsf{r}}^{(8)}$, $\mathrm{u}_{\mathsf{r}}^{(5)}$, $\mathrm{u}_{\mathsf{r}}^{(12)}$, $\mathrm{u}_{\mathsf{r}}^{(14)}$, and $\mathrm{u}_{\mathsf{r}}^{(17)}$.}
\label{tablesu8}
\end{table}

\begin{table}[!ht]
\centering
\renewcommand{\arraystretch}{2.3}
\resizebox{\textwidth}{!}{\small
\begin{tabular}{p{1cm}||p{1.5cm}|p{1.2cm}|p{1.2cm}|p{1.2cm}|p{1.2cm}|p{1.2cm}|p{1.2cm}|p{1.2cm}|}
\hline \hline
$G$ & $\mathsf{r}$ & $\mathrm{u}_{\mathsf{r}}^{(1)}$ & \centering $\mathrm{u}_{\mathsf{r}}^{(2)}$ & \centering $\mathrm{u}_{\mathsf{r}}^{(3)}$ & \centering $\mathrm{u}_{\mathsf{r}}^{(4)}$ & \centering $\mathrm{u}_{\mathsf{r}}^{(5)}$ & \centering $\mathrm{u}_{\mathsf{r}}^{(6)}$ & \hspace{0.25cm} $\mathrm{u}_{\mathsf{r}}^{(7)}$ \\ \hline 
\multirow{13}{*}{$SU(7)$}& $\sf{7}, \overline{7}$ & $1$ & $-$ & $-$ & $-$ & $1$ & $-$ & $1$ \\
                      & $\sf{21}, \overline{21}$ & $-25$ & $15$ & $-$ & $10$ & $-1$ & $3$ & $5$ \\
                      & $\sf{28}, \overline{28}$ & $39$ & $15$ & $-$ & $10$ & $15$ & $3$ & $9$ \\
                      & $\sf{35}, \overline{35}$ & $40$ & $-45$ & $15$ & $-10$ & $-8$ & $9$ & $10$ \\
                      & $\sf{48}$ & $14$ & $30$ & $-$ & $-20$ & $14$ & $6$ & $14$ \\
                      & $\sf{84}, \overline{84}$ & $495$ & $255$ & $15$ & $150$ & $111$ & $33$ & $45$ \\
                      & $\sf{112}, \overline{112}$ & $-194$ & $210$ & $30$ & $140$ & $22$ & $42$ & $46$ \\
                      & $\sf{140}, \overline{140}$ & $-155$ & $165$ & $45$ & $10$ & $13$ & $51$ & $55$ \\
                      & $\sf{189}, \overline{189}$ & $300$ & $465$ & $45$ & $-150$ & $132$ & $75$ & $90$ \\
                      & $\sf{196}, \overline{196}$ & $-1365$ & $765$ & $210$ & $630$ & $-21$ & $153$ & $105$ \\
                      & $\sf{210}, \overline{210}$ & $275$ & $-240$ & $225$ & $-20$ & $-13$ & $114$ & $95$ \\
                      & $\sf{210}', \overline{210}'$ & $3705$ & $2160$ & $195$ & $1180$ & $561$ & $198$ & $165$ \\
                      & $\sf{224}, \overline{224}$ & $340$ & $-300$ & $240$ & $-120$ & $-20$ & $120$ & $100$ \\
                      \hline
\end{tabular}
}
\caption{Representations and their trace-identity indices for $A_6$. In the main text, they are identified with $\mathrm{u}_{\mathsf{r}}^{(1)}$, $\mathrm{u}_{\mathsf{r}}^{(3)}$, $\mathrm{u}_{\mathsf{r}}^{(8)}$, $\mathrm{u}_{\mathsf{r}}^{(5)}$, $\mathrm{u}_{\mathsf{r}}^{(12)}$, $\mathrm{u}_{\mathsf{r}}^{(14)}$, and $\mathrm{u}_{\mathsf{r}}^{(17)}$.}
\label{tablesu7}
\end{table}

\begin{table}[!ht]
\centering
\renewcommand{\arraystretch}{2.2}
\resizebox{\textwidth}{!}{\small
\begin{tabular}{p{1cm}||p{1.5cm}|p{1.2cm}|p{1.2cm}|p{1.2cm}|p{1.2cm}|p{1.2cm}|p{1.2cm}|p{1.2cm}|}
\hline \hline
$G$ & $\mathsf{r}$ & $\mathrm{u}_{\mathsf{r}}^{(1)}$ & \centering $\mathrm{u}_{\mathsf{r}}^{(2)}$ & \centering $\mathrm{u}_{\mathsf{r}}^{(3)}$ & \centering $\mathrm{u}_{\mathsf{r}}^{(4)}$ & \centering $\mathrm{u}_{\mathsf{r}}^{(5)}$ & \centering $\mathrm{u}_{\mathsf{r}}^{(6)}$ & \hspace{0.25cm} $\mathrm{u}_{\mathsf{r}}^{(7)}$ \\ \hline 
\multirow{16}{*}{$SU(6)$}& $\sf{6}, \overline{6}$ & $1$ & $-$ & $-$ & $-$ & $1$ & $-$ & $1$ \\
                      & $\sf{15}, \overline{15}$ & $-26$ & $15$ & $-$ & $10$ & $-2$ & $3$ & $4$ \\
                      & $\sf{20}$ & $66$ & $-60$ & $15$ & $-20$ & $-6$ & $6$ & $6$ \\
                      & $\sf{21}, \overline{21}$ & $38$ & $15$ & $-$ & $10$ & $14$ & $3$ & $8$ \\
                      & $\sf{35}$ & $12$ & $30$ & $-$ & $-20$ & $12$ & $6$ & $12$ \\
                      & $\sf{56}, \overline{56}$ & $456$ & $240$ & $15$ & $140$ & $96$ & $30$ & $36$ \\
                      & $\sf{70}, \overline{70}$ & $-207$ & $180$ & $30$ & $120$ & $9$ & $36$ & $33$ \\
                      & $\sf{84}, \overline{84}$ & $-142$ & $120$ & $45$ & $20$ & $2$ & $42$ & $38$ \\
                      & $\sf{105}, \overline{105}$ & $442$ & $-375$ & $180$ & $-130$ & $-14$ & $69$ & $52$ \\
                      &  $\sf{105}', \overline{105}'$ & $-1196$ & $570$ & $180$ & $500$ & $-44$ & $114$ & $64$ \\
                      & $\sf{120}, \overline{120}$ & $248$ & $420$ & $45$ & $-140$ & $104$ & $66$ & $68$ \\
                      & $\sf{126}, \overline{126}$ & $3210$ & $1905$ & $180$ & $1030$ & $450$ & $165$ & $120$ \\
                      & $\sf{175}$ & $3420$ & $-3690$ & $1320$ & $-1020$ & $-180$ & $270$ & $120$ \\
                      & $\sf{189}$ & $-792$ & $180$ & $360$ & $240$ & $-72$ & $180$ & $108$ \\
                      & $\sf{210}, \overline{210}$ & $611$ & $-1260$ & $810$ & $-120$ & $-133$ & $252$ & $131$ \\
                      &  $\sf{210}', \overline{210}'$ & $-418$ & $1515$ & $360$ & $890$ & $182$ & $231$ & $152$ \\
                      \hline
\end{tabular}
}
\caption{Representations and their trace-identity indices for $A_5$. In the main text, they are identified with $\mathrm{u}_{\mathsf{r}}^{(1)}$, $\mathrm{u}_{\mathsf{r}}^{(3)}$, $\mathrm{u}_{\mathsf{r}}^{(8)}$, $\mathrm{u}_{\mathsf{r}}^{(5)}$, $\mathrm{u}_{\mathsf{r}}^{(12)}$, $\mathrm{u}_{\mathsf{r}}^{(14)}$, and $\mathrm{u}_{\mathsf{r}}^{(17)}$.}
\label{tablesu6}
\end{table}

\clearpage

\bibliographystyle{JHEP}
\bibliography{4comma0.bib}

\end{document}